\documentclass[usenatbib]{mnras}
\usepackage{mathptmx}
\usepackage{color}
\usepackage{xcolor}
\usepackage{graphics,graphicx,amssymb,amsmath}
\usepackage{subfigure}
\usepackage{multirow}
\usepackage{soul}
\usepackage{csvsimple}

\soulregister\citep7
\soulregister\citet7
\soulregister\ref7

\def\cygx{{Cygnus X}}
\def\fc{{\sl FORCAST}}
\def\sf{{\sl SOFIA}}
\def\sp{{\sl Spitzer}}

\usepackage{ulem}

\date{Accepted XXX. Received YYY; in original form ZZZ}

\pubyear{2021}

\begin{document}
\title[]{Completing the Protostellar Luminosity Function in Cygnus-X with SOFIA/FORCAST Imaging}

\author[]{Yingjie Cheng$^{1}$ \thanks{Contact e-mail:yingjiecheng@umass.edu}, Robert A. Gutermuth$^{1}$, Stella Offner$^{2}$, Mark Hemeon-Heyer$^{1}$, Hans Zinnecker$^{3}$ \newauthor{S. Thomas Megeath$^{4}$, Riwaj Pokhrel$^{4}$} \\$^{1}$Department of Astronomy, University of Massachusetts, Amherst, MA, USA \\$^{2}$Department of Astronomy, University of Texas at Austin, Austin, TX, USA \\$^{3}$Universidad Autonoma de Chile, Providencia, Santiago de Chile, Chile \\$^{4}$Ritter Astrophysical Research Center, Dept of Physics and Astronomy, University of Toledo, OH, USA}

\maketitle
\begin{abstract}
We present a new \sf/\fc\ mid-IR survey of luminous protostars and crowded star-forming environments in \cygx, the nearest million-solar mass molecular cloud complex. We derive bolometric luminosities for over 1000 sources in the region with these new data in combination with extant \sp\ and UKIDSS photometry, with 63 new luminous protostar candidates identified by way of the high quality \sf/\fc\ data. By including \fc\ data, we construct protostellar luminosity functions (PLFs) with improved completeness at the high luminosity end. The PLFs are well described by a power law function with an index of $\sim-0.5$. Based on the Herschel temperature and column density measurements, we find no obvious dependence of the PLFs on the local gas temperature, but PLFs in regions of high stellar density or gas column density exhibit some excess at higher luminosities. Through the comparison between our observed PLFs and existing accretion models, both the turbulent core (TC) and the competitive accretion (CA) models are consistent with our results, while the isothermal sphere (IS) model is disfavored. The implications of these results on the star formation process are discussed.

\end{abstract}

%%%%%%%%%%%%%%%%%%%%%%%%%%%%%%%%%%%%%%%%%%%%%%%%%%%%%%%%%%%%%%%%%%%%%%%%%%%%%%
%%%%%%%%%%%%%%%%%%%%%%%%%%%%%%%%%%%%%%%%%%%%%%%%%%%%%%%%%%%%%%%%%%%%%%%%%%%%%%

\section{INTRODUCTION}
\label{s:intro}
Large Infrared (IR) surveys of molecular clouds have consistently shown that low-mass protostars are less luminous than expected, which is well known as the 'luminosity problem' and challenges traditional star formation theories \citep{kenyon1990, dunham2014}. A variety of theoretical ideas have been proposed to explain this phenomenon, including episodic accretion, mass-dependent accretion, variable envelope infall, and accelerating star-formation (e.g. \citealt{offner2011, dunham2012, padoan2014, fischer2017}). The solution to the luminosity problem would directly shed light on some fundamental problems of star formation (e.g. the time required to form a low-mass star) \citep{mckee2011}. Different star formation theories predict different prototellar mass functions (PMF) and protostellar luminosity functions (PLF). The PMF represents the present-day mass distribution of protostars \citep{mckee2010}. Since it depends on the stellar IMF and the protostellar accretion histories, it can be used to infer the nature of the star formation process. However, the PMF is currently inaccessible to direct observation. Masses have been inferred for a small number of protostars using disk rotation (e.g. \citealt{tobin2012}), but no statistical sample of protostellar masses exists. As a consequence, the best way to constrain the PMF currently is through observations of the PLF, the present-day luminosity function of protostars. The observed PLF, in combination with models for accretion and protostellar evolution, can be used to discriminate between different star formation models. Thus, the PLF provides a powerful tool for addressing the ’luminosity problem’ and in turn testing existing star formation theories.

The scope of the problem hinges critically on obtaining a census of young stellar objects (YSOs) and extending their spectral energy distribution (SED) coverage through mid-IR wavelengths (\citealt{Kryu2012}, K12 afterwards). At a distance of $\sim 1.4$ kpc, the \cygx\ complex is one of the most active nearby star-formation regions \citep{rygl2012}. The huge sample of protostars in \cygx\ as well as the wide range of star-forming environments it encompasses set the stage for completing such census and photometry of YSOs. The SEDs of identified protostars help characterize the aggregate PLF, as well as how the PLF is affected by natal environment.

The \cygx\ \sp\ Legacy Survey mapped 24 sq. deg. of the \cygx\ complex with the IRAC and MIPS instruments on board the Spitzer Space Telescope (\citealt{beerer2010, Kryu2014}, K14 afterwards). The latest \sp\ data and catalogue have provided a census of protostars in \cygx\ of unprecedented richness (\citealt{pokhrel2020}, Gutermuth et al. in prep). Over 21,800 YSO candidates with excess IR emission were identified, among which over 2,400 are young embedded protostars, distinguished by flat/rising mid-IR SEDs that indicate the presence of cold, dusty spherical envelopes. However, the capacity of \sp\ to constrain the luminosity of bright sources is largely inhibited by confusion and saturation in the MIPS 24~$\mu$m data, especially in several bright, clustered star-forming regions. Although the Wide-field Infrared Survey Explorer (WISE) and the Midcourse Space Experiment (MSX) have coverage in \cygx\ at similar wavelengths, the corresponding archival data suffers from low angular resolution leading to source confusion. 

In order to obtain the luminosities of bright, closely clustered protostars and in turn complete the PLF coverage for sources with bolometric luminosities $L_{\rm bol}>10 L_{\odot}$, we have conducted new observations with the Faint Object infraRed CAmera for the \sf\ Telescope (\fc). \fc\ is a dual-channel mid-infrared camera and spectrograph sensitive to wavelengths between 5-40~$\mu$m, while each channel has an instantaneous field-of-view of $3.4^{\prime}\times3.2^{\prime}$. By combining bright protostars identified by \fc\ with the existing \sp\ point source catalog, the coverage and luminosity estimates of bright protostars are greatly improved. Furthermore, testable differences among the PLF model predictions are located exactly in the $10<L_{\rm bol}<1000L_{\odot}$ range \citep{offner2011}, including the \fc\ imaging survey will make the comparison between observations and theories more feasible.
The enclosed wavelength bands have a profound effect on the estimation of SED slopes and luminosities of protostars (e.g., \citealt{hops2016}). In order to guarantee a robust PLF measurement, we limited the sample to sources with complete photometry. For sources without MIPS 24~$\mu$m detections, \fc\ provides better limits on luminosity in regions of high source confusion.

In \cygx, the large sample of protostars enables us not only to characterize the PLF, but also to probe how the PLF varies with different local environments. To test whether the PLF is biased to higher luminosities in dense regions as found for Orion by \citealt{kryu2012}, \citealt{kryu2014} compared the PLF in regions of high and low YSO density within \cygx. They argued that the PLF in \cygx\ varies within molecular complexes and depends on the local environment in which protostars form. These results motivate us to further compare ensembles of protostars in different environments, with the goal of understanding how environmental factors influence the gas dynamics and the eventual luminosity distribution of the protostars. We divide the total protostar sample into several subsets according to the local gas temperature, gas column density, and proximity of neighboring protostars with sufficient sample sizes to constrain PLF shapes among the subsets. Such analysis further allows us to test the predictions of competing theoretical models. Throughout the paper, the distance to \cygx\ is assumed to be 1.4 kpc from the maser parallax study \citep{rygl2012}.

The rest of the paper is organized as follows: In Section \ref{s:obs}, we describe the data sources and preliminary data reduction processes. The calculation of protostar luminosities or upper limits, and the construction of PLFs are given in Section \ref{s:res}. In Section \ref{s:dis}, we discuss how our results compare with existing accretion models and some indications on the star formation theories. Our main conclusions are summarized in Section \ref{s:con}.

%%%%%%%%%%%%%%%%%%%%%%%%%%%%%%%%%%%%%%%%%%%%%%%%%%%%%%%%%%%%%%%%%%%%%%%%%%%%%%
%%%%%%%%%%%%%%%%%%%%%%%%%%%%%%%%%%%%%%%%%%%%%%%%%%%%%%%%%%%%%%%%%%%%%%%%%%%%%%

\section{Observation and data reduction}
\label{s:obs}

\subsection{Spitzer SESNA Catalogue}
\label{ss:sesna}
A full redevelopment of the \cygx\ \sp\ Legacy Survey has been produced as part of the \sp\ Extended Solar Neighborhood Archive (SESNA). SESNA is a collection of uniformly produced \sp\ mosaics, source catalogs and YSO identifications with corresponding completeness maps (\citealt{pokhrel2020}, Gutermuth et al. in prep). The full archive spans 92 sq. deg. (plus another 16 sq. deg. of extragalactic fields to explore residual contamination rates), which includes 24 sq. deg. of \cygx\ coverage. Within \cygx, over ten million sources have been detected, among which $\sim21$ thousand have been identified as YSOs.

The SESNA protostar catalogue and photometry in \cygx\ serve as our main protostar sample. Additionally, \fc\ data improve the protostellar sampling at high luminosities as well as the completeness limits. For each source, the SESNA catalogue lists the sky coordinates in RA and DEC, magnitudes at \sp\ IRAC 4 bands (3.6~$\mu$m, 4.5~$\mu$m, 5.8~$\mu$m, 8.0~$\mu$m) and the MIPS 24 band (24.0~$\mu$m), and also the classifications. We will further use this information to identify protostars and build their SEDs in the following sections.

\subsection{SOFIA/FORCAST Data}
\label{ss:data}

The capacity of the \sp\ data to constrain the luminosity of bright protostars is inhibited by confusion and saturation, especially in bright and clustered star-forming areas. Therefore, \sf/\fc\ imaging is essential for obtaining the luminosity of bright, closely clustered protostars in \cygx\ to complete the PLF coverage for the higher luminosity end. For \fc\ observations, the field selection is based on SESNA protostars that were missing MIPS 24~$\mu$m photometry and had poor 24~$\mu$m completeness or no 24~$\mu$m coverage, so that sources with rising SEDs and with a bolometric luminosity of $L_{\rm bol} > 10 L_{\odot}$ could be included. We tuned the integration depths to reach this constraint of $L_{\rm bol}$.

The observations are conducted by \fc\ during its Cycle~5, Cycle~7 and Cycle~8 in 2017, 2019 and 2021 respectively (Program ID: 05\_0181, 07\_0225, 08\_0181; PI: R. Gutermuth). Apart from these new observations, we have also enclosed two archival \fc\ observations in \cygx, DR7 (De Buizer et al.) and DR21 OH central (Hill et al.). An astrometry adjustment was made to the DR7 observation fields to ensure that the sources can be accurately matched to SESNA and other catalogs.

Our observations of \cygx\ mainly cover the 19.7~$\mu$m and 31.5~$\mu$m \fc\ bands, where the spatial resolution is $\sim4^{\prime\prime}$. After data calibration, we used the IDL-based interactive photometry visualization tool PhotVis \citep{guter2008} to identify all point sources detected and perform aperture photometry at each position. For object detections, we set the FWHM to be 3.1$^{\prime\prime}$ (4 pixels) and the sigma threshold to be 10 (the number of standard deviations above local background noise for valid auto-detections), and then visually filtered out spurious detections. The aperture radius was selected to be 3.8$^{\prime\prime}$ (5 pixels) for both wavelengths, while the inner and outer sky annulus radii were 7.7$^{\prime\prime}$ and 15.4$^{\prime\prime}$ (10-20 pixels) respectively. Since the image units were already flux-calibrated, we could simply de-convert the resulting magnitudes to fluxes in $Jy$ at each wavelength. We have further verified the \fc\ calibration against the \sp\ data by checking the SED slope correlations (see Fig \ref{f:alpha}).

Due to instrumental problems, Cycle~5 observations show double-beam features in most fields, where a single source is detected twice at a close distance. To calibrate those double-beam detections, we reduced the aperture size to get the photometry for both beams, added their fluxes and applied a scaling factor to determine the real flux of that single source. In particular, the FWHM was set to be 3.8$^{\prime\prime}$ and the sigma threshold was 8. The aperture radius became 2.3$^{\prime\prime}$ while the inner and outer sky annulus radius was unchanged. Since two of the Cycle~5 fields that suffer double-beaming have been re-observed with decent data quality in Cycle~8, the scaling factor was derived by comparing the photometry of the same sources using the same aperture and annulus sizes.

Finally, a total of 78 sources (12 cycle~5 + 43 cycle~7 + 8 cycle~8 + 15 archival) are detected by \fc, among which 55 can be matched (distance $<~2^{\prime \prime}$) to the SESNA catalogue (Table \ref{tab:fc}).

\begin{table*}

\caption{\fc\ detected sources matched to the SESNA catalogue.}
\begin{filecontents*}{forcast.csv}
OBSID,RA,DEC,SESNA~Name,Flux_{19},Flux_{31},Class
(cycle-num),(J2000),(J2000),,(Jy),(Jy),
c8-130-162,308.1485,40.2683,J203235.54+401605.9,3.52\pm0.07,,\rm II
c8-163-194,305.4496,37.5048,J202147.96+373017.4,1.45\pm0.07,8.24\pm0.07,\rm I
c8-215-253,306.824,37.3729,J202717.77+372222.9,1.04\pm-0.07,2.22\pm0.07,\rm U
c8-029-048,305.8638,37.5817,J202327.31+373453.9,2.32\pm0.05,6.71\pm0.06,\rm I
,305.8488,37.5931,J202323.71+373535.3,0.57\pm0.05,7.43\pm0.06,\rm I
,305.842,37.5857,J202322.09+373508.4,0.26\pm0.05,1.97\pm0.06,\rm I
,306.8468,37.3903,J202723.30+372326.2,2.28\pm0.07,8.65\pm0.07,\rm I
c7-019-057,305.1638,39.6327,J202039.32+393757.6,17.77\pm0.37,78.00\pm1.70,\rm I
c7-073-138,306.6348,39.9558,J202632.35+395720.8,3.12\pm0.13,12.37\pm0.15,\rm I
c7-077-132,308.119,40.3282,J203228.56+401941.6,3.55\pm0.15,5.53\pm0.08,\rm I
,308.0858,40.3306,J203220.60+401950.2,0.41\pm0.04,1.71\pm0.05,\rm I
,308.092,40.3381,J203222.08+402017.1,2.49\pm0.11,6.00\pm0.09,\rm I
,308.088,40.3405,J203520.65+402017.1,0.35\pm0.04,1.55\pm0.05,\rm I
c7-086-139,309.7612,42.4145,J203902.69+422452.2,13.96\pm0.58,2.79\pm0.06,\rm I
,309.7583,42.4165,J203901.99+422459.3,5.65\pm0.24,72.82\pm0.84,\rm I
,309.7655,42.4249,J203903.72+422529.7,0.59\pm0.04,20.74\pm0.24,\rm I
,309.767,42.4281,J203904.07+422541.0,0.18\pm0.04,3.13\pm0.06,\rm I
,309.7624,42.4311,J203902.97+422552.0,1.69\pm0.08,2.62\pm0.06,\rm I^*
,309.7669,42.4355,J203904.07+422607.8,0.76\pm0.05,0.81\pm0.05,\rm I
c7-089-168,308.4988,41.3762,J203359.71+412234.4,0.52\pm0.04,4.21\pm0.07,\rm U
c7-133-184,309.6715,42.6564,J203841.16+423923.1,47.12\pm1.94,0.57\pm0.05,\rm II
,309.6554,42.6591,J203837.29+423932.9,3.61\pm0.15,5.85\pm0.08,\rm I^*
,309.6673,42.6649,J203840.16+423953.6,17.65\pm0.73,4.00\pm0.07,\rm II
c7-139-195,309.0137,41.661,J203603.28+413939.5,0.58\pm0.04,0.91\pm0.05,\rm I
,309.0302,41.6647,J203607.25+413952.8,23.04\pm0.95,90.29\pm1.04,\rm II
,309.0147,41.6682,J203603.54+414005.5,0.57\pm0.04,11.14\pm0.14,\rm II
,309.0314,41.6692,J203607.53+414009.0,0.66\pm0.04,29.99\pm0.35,\rm II
c7-140-177,309.743,42.3162,J203858.31+421858.3,0.46\pm0.04,2.54\pm0.08,\rm I
c7-144-158,307.8152,40.0404,J203115.65+400225.3,0.74\pm0.06,0.16\pm0.05,\rm U
,307.7865,40.0465,J203108.77+400247.5,1.09\pm0.06,0.44\pm0.05,\rm I
,307.7969,40.0521,J203111.25+400307.6,0.43\pm0.04,73.61\pm0.85,\rm II
,307.793,40.0546,J203110.31+400316.5,5.56\pm0.23,35.23\pm0.41,\rm I
,307.8038,40.0565,J203112.92+400323.3,64.91,3.47\pm0.06,\rm I^*
c7-159-231,307.6226,41.266,J203029.42+411557.7,9.51,3.03\pm0.06,\rm I
,307.6198,41.2664,J203028.76+411559.1,10.25,5.78\pm0.08,\rm I
c7-162-231,306.0138,37.61,J202403.32+373636.0,32.53,2.69\pm0.06,\rm U
,306.0118,37.6103,J202402.82+373636.9,0.40\pm0.04,3.65\pm0.06,\rm I
,305.9837,37.6289,J202356.10+373744.1,6.14\pm0.31,1.71\pm0.05,\rm I
c7-232-248,310.8678,42.8167,J204328.26+424900.1,27.64\pm1.39,9.34\pm0.21,\rm I
c7-250-256,310.6384,42.9471,J204233.21+425649.6,2.28\pm0.12,192.79\pm4.20,\rm U
c5-123-169,305.4232,37.435,J202141.57+372606.0,1.26\pm0.07,74.2,\rm I
c5-211-258,305.4354,37.4439,J202144.50+372637.9,4.34\pm0.22,172.15,\rm I
c5-277-291,309.6474,42.6205,J203835.38+423713.7,6.40\pm0.34,56.1,\rm I
a-192-203,307.0231,40.8723,J202805.54+405220.2,,16.26\pm0.84,\rm II
,307.0669,40.8818,J202816.06+405254.6,,13.32\pm0.69,\rm I
,307.0426,40.8938,J202810.22+405337.7,,2.56\pm0.13,\rm I
a-204-214,307.0188,40.876,J202804.63+405234.2,0.4,14.56\pm0.76,\rm U
a-090-109,309.7591,42.3664,J203902.17+422159.1,6.14,0.37\pm0.02,\rm U
,309.7383,42.3781,J203857.19+422241.1,27.64,13.32,\rm I
,309.7403,42.3801,J203857.67+422248.5,2.28,2.56,\rm U
,309.7542,42.3806,J203901.01+422250.2,1.26,,\rm U
,309.7519,42.4102,J203900.45+422436.7,4.34,,\rm I
,309.7583,42.4165,J203901.99+422459.3,6.4,,\rm I
,309.7507,42.4102,J203900.16+422436.8,,14.56,\rm U
,309.7612,42.4145,J203902.69+422452.2,,0.37,\rm I
\end{filecontents*}
\csvreader[no head,
  tabular=lllllll,
  table head=\hline,
  late after last line=\\\hline]{forcast.csv}
  {1=\one,2=\two,3=\three,4=\four,5=\five,6=\six,7=\seven}
  {$\one$ & $\two$ & $\three$ & $\four$ & $\five$ & $\six$ & $\seven$}
\label{tab:fc}

\vspace{0.2cm}
Note: The OBS ID is in the format of cycle number(or archival)-observation number, and the position is in RA and DEC. Fluxes and flux uncertainties in $Jy$ are listed for both 19.7~$\mu$m and 31.5~$\mu$m bands. The corresponding SESNA source indexes and classifications are shown in the last two columns. Class '$\rm I^*$' refers to deeply embedded protostars (classifying these sources as bona fide Class 0 protostars requires more detailed treatment). Class II refers to pre-main sequence stars with disks, a few protostars can hide here, but the fraction is low. Class U means unclassified sources, which are usually field stars but could hide others at low concentration.

\end{table*}

%%%%%%%%%%%%%%%%%%%%%%%%%%%%%%%%%%%%%%%%%%%%%%%%%%%%%%%%%%%%%%%%%%%%%%%%%%%%%%

\subsection{Additional Archival Datasets}
\label{ss:arx}

When neither MIPS 24~$\mu$m data nor \fc\ data are available, the 22~$\mu$m flux taken from the Wide-field Infrared Survey Explorer (WISE) was used to constrain the source SED at high wavelength. From existing WISE observations in \cygx, we are able to derive a source list and match it to the SESNA YSO catalogue. Among protostars with valid MIPS 24~$\mu$m detections, 74\% also have WISE 22~$\mu$m data, and the 22~$\mu$m fluxes are well-correlated with the 24~$\mu$m fluxes or completeness. However, although the WISE 22~$\mu$m detections are less sensitive than the MIPS 24~$\mu$m detections, a lot more sources were identified by WISE. This big discrepancy of source counts indicates that a reasonable amount of false sources exist in the WISE data. Moreover, the big beam of WISE frequently leads to overestimation of the 22~$\mu$m fluxes of real sources it does find (e.g. \citealt{guter2015}). Therefore, we only use WISE data to derive the upper limits of the protostar luminosity, especially when the MIPS 24~$\mu$m completeness is unavailable.

In addition, the Herschel Orion Protostar Survey (HOPS) \citep{hops2016} data are adopted for testing the $L_{\rm MIR}$-to-$L_{\rm bol}$ relation for luminosity calculations in Section \ref{ss:method}. HOPS is a sample of 410 YSOs in the Orion molecular clouds selected from \sp\ data. Herschel PACS observations at 70, 100, and 160~$\mu$m yielded far-infrared photometric data points. With near-infrared photometry from the Two Micron All Sky Survey (2MASS), mid and far-infrared data from \sp\ and Herschel, plus sub-millimeter photometry from APEX, their SEDs cover 1.2-870~$\mu$m and sample the peak of the protostellar envelope emission at $\sim$100~$\mu$m.

Herschel observations have also provided the temperature map and gas column density map in \cygx, which we use to study the environmental dependence of PLFs in Section \ref{s:dis}. Herschel’s unprecedented angular resolution and sensitivity in the far-infrared enables dust emission maps of superb quality over large areas of sky. \citealt{pokhrel2020} used thermal dust emission from Herschel maps and performed modified blackbody fits in three Herschel wave bands to derive the column density and temperature maps. Here we use dust temperature as a proxy for gas temperature in \cygx.

%%%%%%%%%%%%%%%%%%%%%%%%%%%%%%%%%%%%%%%%%%%%%%%%%%%%%%%%%%%%%%%%%%%%%%%%%%%%%%

\subsection{Data Synthesis}

By requiring a distance $<~2^{\prime \prime}$, we were able to match \fc\ detected sources to the full SESNA catalogue and the WISE catalogue. Although the full SESNA catalog was used for matching, we mainly work with the protostellar types (Class 0 and I) hereafter, since we are interested in the luminosities of protostars. For matched sources, we derived source fluxes from the 2MASS J, H and Ks bands, \sp\ IRAC 3.6, 4.5, 5.8 and 8.0~$\mu$m bands and MIPS 24~$\mu$m band, WISE 22~$\mu$m band (only for upper limits) and also \fc\ 19 \& 31~$\mu$m bands, and in turn build a relatively complete SED in the near-to-mid IR regime for each protostar candidate.

We obtain the flux upper limits from the \fc\ noise maps for any SESNA protostars that lack \fc\ photometry but were covered in the field of view. By conducting the same aperture photometry to the corresponding noise map, the $1\sigma$ flux uncertainty of each source was obtained, and then the $5\sigma$ flux uncertainties were treated as their flux upper limits.

%%%%%%%%%%%%%%%%%%%%%%%%%%%%%%%%%%%%%%%%%%%%%%%%%%%%%%%%%%%%%%%%%%%%%%%%%%%
%%%%%%%%%%%%%%%%%%%%%%%%%%%%%%%%%%%%%%%%%%%%%%%%%%%%%%%%%%%%%%%%%%%%%%%%%%%

\section{Analysis and Results}
\label{s:res}

\subsection{Determination of Bolometric Luminosity} \label{ss:method}
We compute the bolometric luminosities of protostars using a technique first described in \citealt{kryu2012}. They developed the method empirically using the mid-IR spectral index and mid-infrared luminosities of protostars in the \sp\ c2d legacy program with known bolometric luminosities. Thus, the calculation of the SED slope $\alpha$ is essential to constrain the bolometric luminosities.

The mid-IR SED has been adopted to identify and classify YSOs since the IRAS (Infrared Astronomical Satellite) era. \citealt{greene1994} succeeded in deriving the SED slope $\alpha$ from the 2.2-20~$\mu$m spectra (using K, L, M, N, and Q bands) and went on to classify YSOs based on the range of $\alpha$. Sources with $\alpha>0.3$ were classified as Class I, $0.3>\alpha \geq-0.3$ as 'Flat spectrum', $-0.3>\alpha \geq-1.6$ as Class II, and $\alpha<-1.6$ as Class III YSOs. However, blind adherence to $\alpha$ in classification can be reddening degenerate \citep{muench2007}. Modern classification techniques are more reddening-independent and more adaptable to varying data availability (e.g. \citealt{guter2009}). While SESNA uses the process described in \citealt{guter2009} to classify YSOs, many sources are missing vital data to constrain $\alpha$ through the mid-IR wavelengths. As noted above, we employ \fc\ data to enhance the number of Spitzer-identified protostars that have the requisite 20-30~$\mu$m photometry for constraining $L_{\rm bol}$ via the  \citealt{kryu2012} technique. As we will show in Section~\ref{ss:yso} below, several sources with valid detections in both \fc\ and \sp/MIPS are available to test the utility of \fc\ observations for this purpose.

To be specific, $\alpha$ is obtained by fitting a line to the $log(\lambda F_{\lambda})$ versus $log(\lambda)$ plot over a certain wavelength range. Here we make use of the four \sp\ IRAC bands (3.6~$\mu$m, 4.5~$\mu$m, 5.8~$\mu$m and 8.0~$\mu$m), the \sp\ MIPS 24 band (24.0~$\mu$m) and also the \fc\ 31 band (31.5~$\mu$m). With one or more missing bands, one can still obtain a fairly good estimation or upper limit for $\alpha$, as we will discuss in Section~ \ref{sss:a}.

The next step is to calculate the mid-infrared luminosities of the protostars. As described in \citealt{kryu2012}, mid-IR luminosities ($L_{\rm MIR}$) can be calculated by integrating the SED over all available fluxes. The equation is shown below, which is a rectangular integration over each band (summing the product of the flux and bandwidth assuming a flat spectrum) and then converting to luminosity:

\begin{equation}
\begin{split}
L_{\mathrm{MIR}}=&\left[19.79 F_{\nu}(J)+16.96 F_{\nu}(H)+10.49 F_{\nu}\left(K_{s}\right)\right.\\ &+5.50 F_{\nu}(3.6)+4.68 F_{\nu}(4.5)+4.01 F_{\nu}(5.8) \\ & \left.+4.31 F_{\nu}(8.0)+0.81 F_{\nu}(24)\right] \times 10^{-6} \times d^{2} L_{\odot} 
\end{split}
\label{e:lmir}
\end{equation}
where $d$ is the distance to the cloud in pc and fluxes $F_{\nu}$ are in Jy.
Finally, we convert the mid-IR luminosities to bolometric luminosities ($L_{\rm bol}$) through the following relationship given by \citealt{kryu2012}:

\begin{equation}
\begin{split}
\frac{L_{\mathrm{MIR}}}{L_{\mathrm{bol}}}&=(-0.466 \pm 0.014 \times \log (\alpha)+0.337 \pm 0.053)^{2}~(\alpha>=0.3) \\
\frac{L_{\mathrm{MIR}}}{L_{\mathrm{bol}}}&=0.338~(\alpha<0.3)
\end{split}
\label{e:lbol}
\end{equation}

\begin{figure}
\includegraphics[width=1\linewidth]{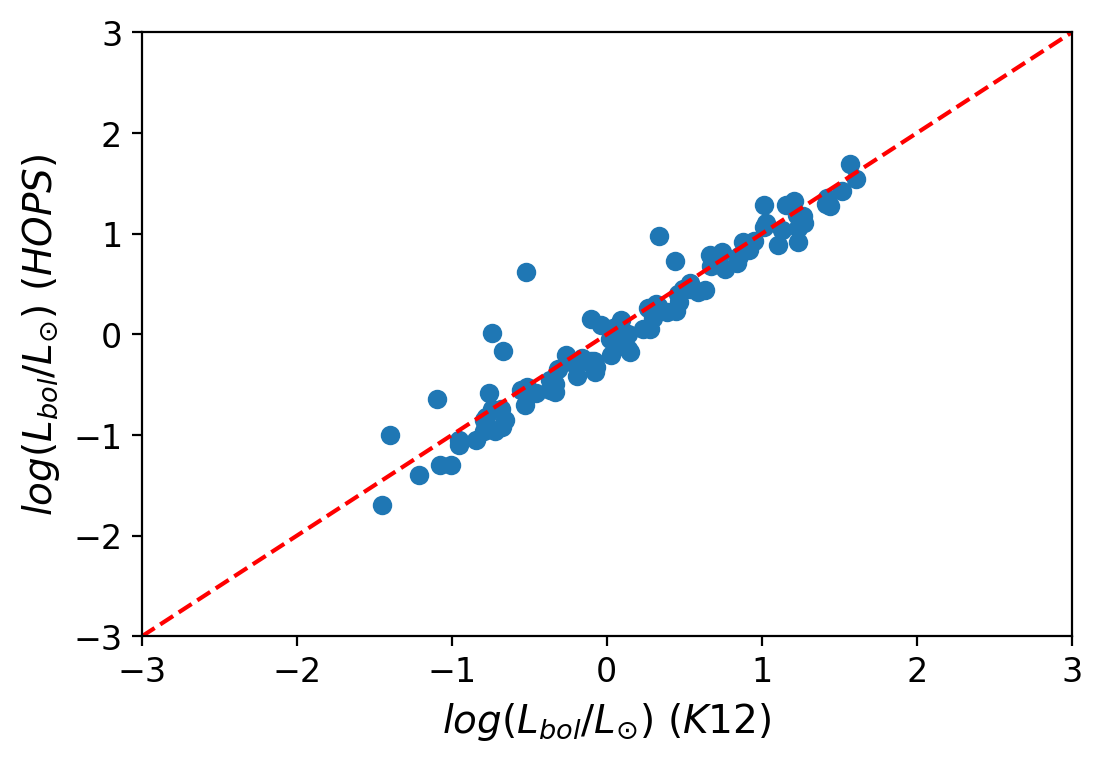}
\caption{$L_{\rm bol}$ directly taken from the HOPS dataset (y axis) versus $L_{\rm bol}$ calculated from Eq. \ref{e:lbol} (x axis).}
\label{f:hops}
\end{figure}

\begin{figure}
\includegraphics[width=1\linewidth]{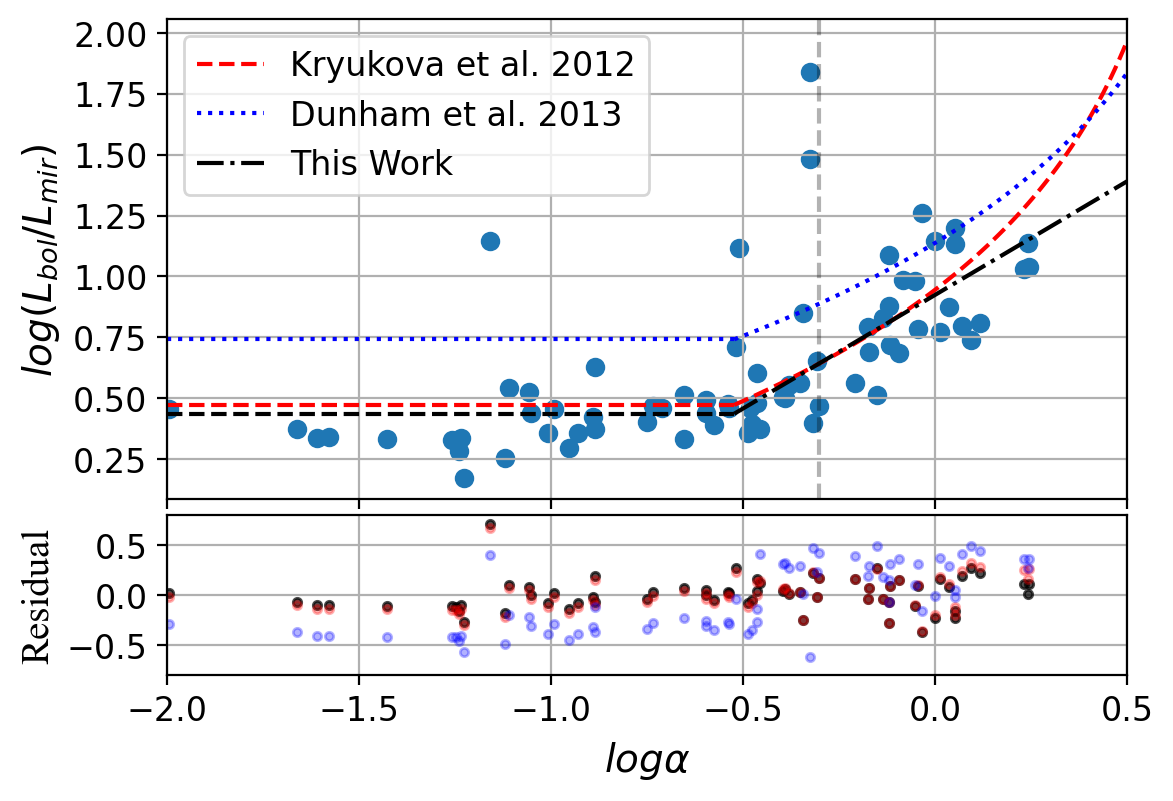}
\caption{The $L_{\rm bol}$-to-$L_{\rm MIR}$ ratio as a function of $\alpha$ in the log-log scale. \citealt{kryu2012}'s fitting function is plotted in red, \citealt{dunham2013}'s function is in blue, and the best linear fit of this work is in black. The residuals of all three fittings are shown in the bottom panel.}
\label{f:hops_fit}
\end{figure}

The ratio between $L_{\rm bol}$ and $L_{\rm MIR}$ is a simple function of the SED slope $\alpha$, so well-constrained $\alpha$ and $L_{\rm MIR}$ are required to provide useful constraints on the $L_{\rm bol}$ of protostars. Before applying this $L_{\rm MIR}$-to-$L_{\rm bol}$ relation (Eq. \ref{e:lbol}) to our study, we test it with the Herschel Orion Protostar Survey (HOPS) \citep{hops2016}. HOPS provides SEDs and model fits of 330 YSOs (predominantly protostars) in the Orion molecular clouds. The dataset directly gives us the measured bolometric luminosities ($L_{\rm bol}$) by sampling the full 1-1000~$\mu$m SED and also the model-derived total luminosities ($L_{tot}$). We also use the mid-IR portion of the HOPS SEDs to compute $\alpha$, $L_{\rm MIR}$, and $L_{\rm bol}$ as described above. Fig \ref{f:hops} compares The $L_{\rm bol}$ taken from the HOPS dataset and those calculated from Kryukova et al.'s equation. Most of the data points lie on the 1-to-1 line (red dashed line), which provides a strong support for the $L_{\rm MIR}$-to-$L_{\rm bol}$ relation given by \citealt{kryu2012}. To further check the function form and parameters in Eq. \ref{e:lbol}, Fig \ref{f:hops_fit} shows the $L_{\rm bol}$-to-$L_{\rm MIR}$ ratio as a function of $log(\alpha)$. The HOPS data points (blue) follow Eq. \ref{e:lbol} (red dashed line) quite well and flatten at $\alpha<0.3$ as expected. By focusing on the range of $\alpha>0.3$, we are able to fit a straight line to the data points:
\begin{equation}
\log \left(L_{\rm bol} / L_{\rm MIR}\right)=0.933 \pm 0.118(\log \alpha)+0.924
\label{e:linear}
\end{equation}
The offset is determined by the mean value at the $\alpha<0.3$ range. 

\citealt{dunham2013} also tried to test Eq. \ref{e:lbol} with a slightly different dataset. By including 12 and 22~$\mu$m photometry from WISE and 350~$\mu$m photometry from SHARC-II and discarding any IRAS photometry, they re-derived \citealt{kryu2012}’s empirical correlation with new calculations of $L_{\rm bol}$.
Our best-fit function is compared with \citealt{kryu2012}'s and \citealt{dunham2013}'s results in the right panel of Fig \ref{f:hops_fit}. The $\chi^{2}$ per degree of freedom of our linear fitting is 1.35, which is only slightly smaller than 1.44 given by Eq. \ref{e:lbol}, but significantly smaller than 3.45 given by \citealt{dunham2013}'s function. Thus, we conclude that Eq. \ref{e:lbol} is sufficiently accurate to characterize the relation between $L_{\rm MIR}$ and $L_{\rm bol}$.

%It is worth mentioning that the bolometric luminosity ($L_{\rm bol}$) isn't equivalent to the total luminosity ($L_{tot}$) for a protostar. By concept, the PLF should reflect the distribution of the total luminosities instead of $L_{\rm bol}$. Usually, to convert $L_{\rm bol}$ to $L_{tot}$ needs a scaling factor of $\sim2$, which has been supported by the HOPS data \citep{fischer2017}.

%%%%%%%%%%%%%%%%%%%%%%%%%%%%%%%%%%%%%%%%%%%%%%%%%%%%%%%%%%%%%%%%%%%%%%%%%%%%%%

\subsection{SESNA+FORCAST Cygnus X YSOs} \label{ss:yso}

\subsubsection{SED Slope Fitting}
\label{sss:a}

The SESNA protostar catalogue is derived from a multiple-phase classification process, where some phases intentionally don't require certain bandpasses which could lose sensitivity in crowded or nebulous areas. However, missing some bandpasses may affect our calculation of $\alpha$ and $L_{\rm MIR}$, and in turn affect the determination of $L_{\rm bol}$. In order to obtain robust constraints on $L_{\rm bol}$ without causing significant biases on the final sample, we need to determine which bandpasses to require in our calculation. As seen in Eq. \ref{e:lmir}, missing bands can obviously contribute to a downward bias on $L_{\rm MIR}$, with an amount depending on which bands are missing and the SED shape of the source. Similarly, the estimation of $\alpha$ can be biased by band selection, which adds to the overall error budget for $L_{\rm bol}$, given Eq. \ref{e:lbol}. Of the $\sim$2000 SESNA protostar candidates, over 900 sources have full wavelength coverage from 3.6~$\mu$m to 24~$\mu$m (IRAC + M24). This large sample of well-detected sources enables us to experiment on the band selection and to better constrain the data requirements. 

\begin{figure}
\includegraphics[width=1\linewidth]{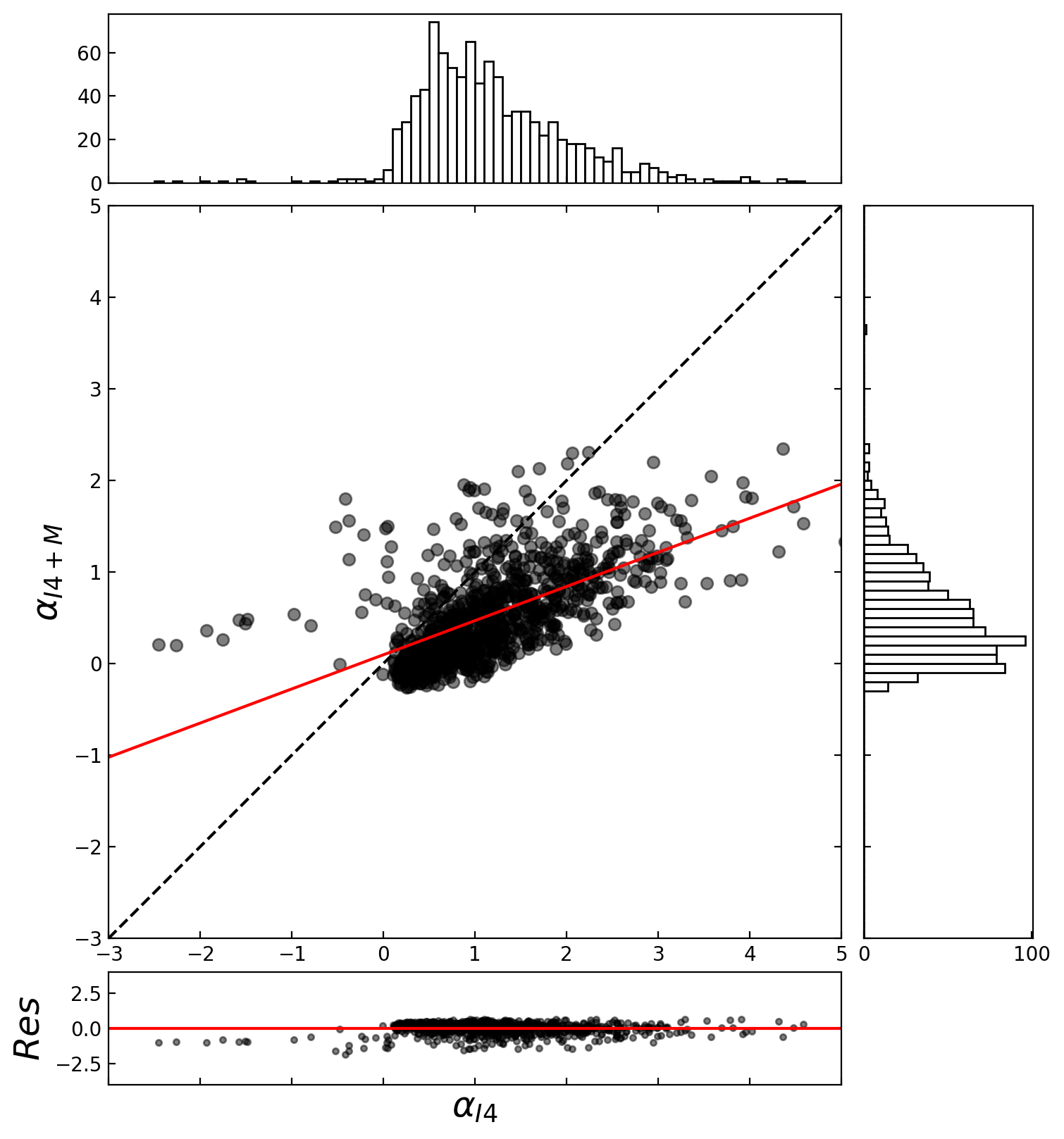}
\caption{Comparing the SED slope fitted by IRAC 4 bands only or by IRAC 4 bands + MIPS 24 band. The black dashed line shows the 1-to-1 relation, and the red solid line shows the best linear fit.}
\label{f:alpha45}
\end{figure}

\citealt{kryu2012} used the four \sp\ IRAC bands and the MIPS 24 band (3.6-24.0~$\mu$m) for SED slope fitting (see also \citealt{muench2007}). Since the wavelengths of the four IRAC bands are relatively close, it is natural to ask that whether long wavelength photometry at $\sim$20-30~$\mu$m is essential to constrain $\alpha$ and $L_{\rm MIR}$. The majority of sources in the SESNA catalogue have complete IRAC photometry, while many luminous sources lack MIPS 24~$\mu$m data. We compare the SED slope fitted by IRAC 4 bands only ($\alpha_{I4}$) with the SED slope fitted by IRAC 4 bands + MIPS 24 band ($\alpha_{I4+M}$) in Fig \ref{f:alpha45}. $\alpha_{I4}$ are typically larger than $\alpha_{I4+M}$ with some scatter, which indicates that the $\sim$20-30~$\mu$m photometry is needed to optimize the SED slope fitting and should always be required in our sample. As expected, \fc\ data provide a good replacement when the MIPS 24 band data are missing. 

\begin{figure}
\includegraphics[width=1\linewidth]{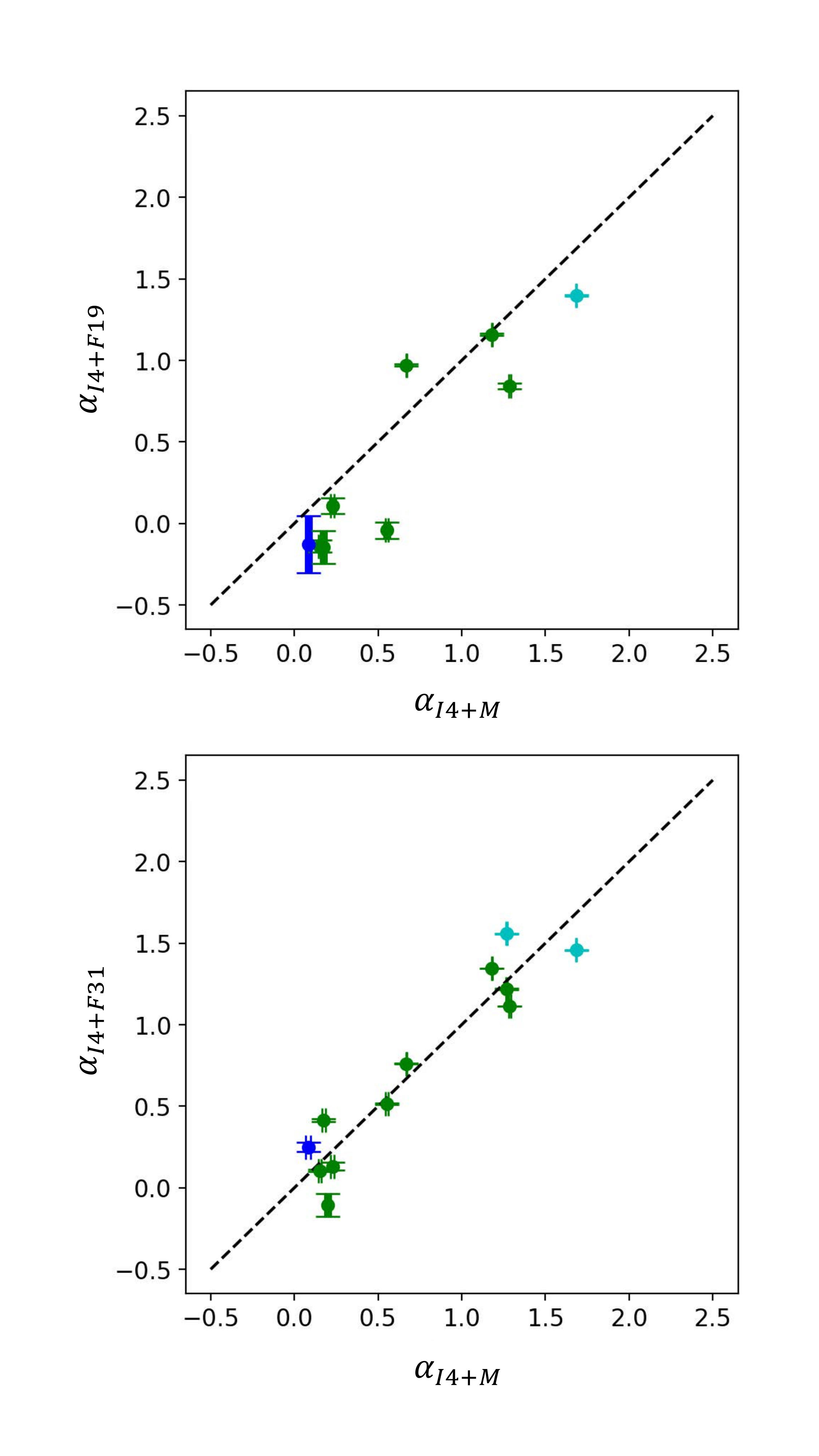}
\caption{SED slope correlations with or without \fc\ data, different classes of sources are marked in different colors (Class $\rm I^*$: cyan; Class I: green; Class II: blue).}
\label{f:alpha}
\end{figure}

\begin{figure*}
\includegraphics[width=1\linewidth]{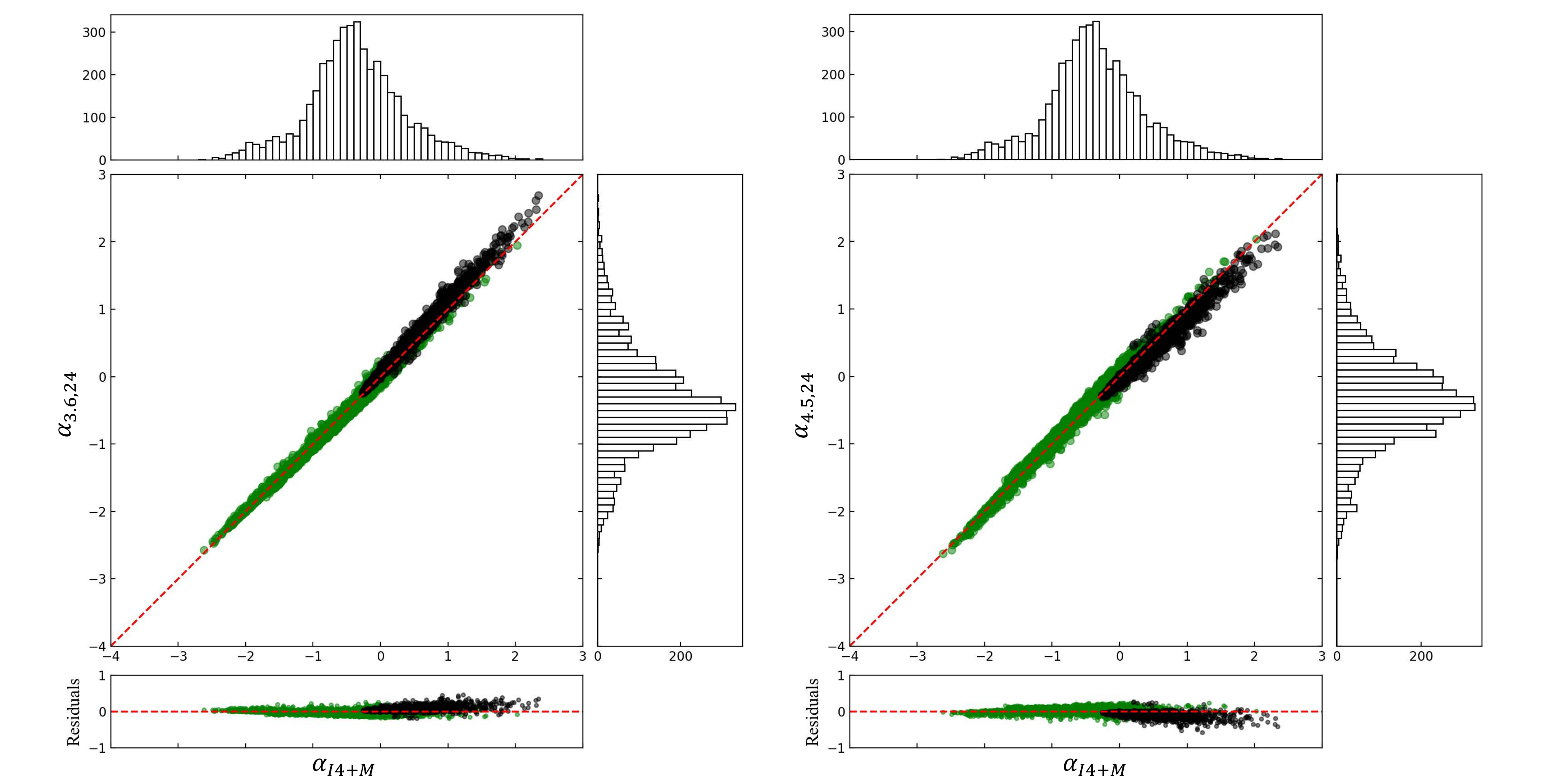}
\includegraphics[width=1\linewidth]{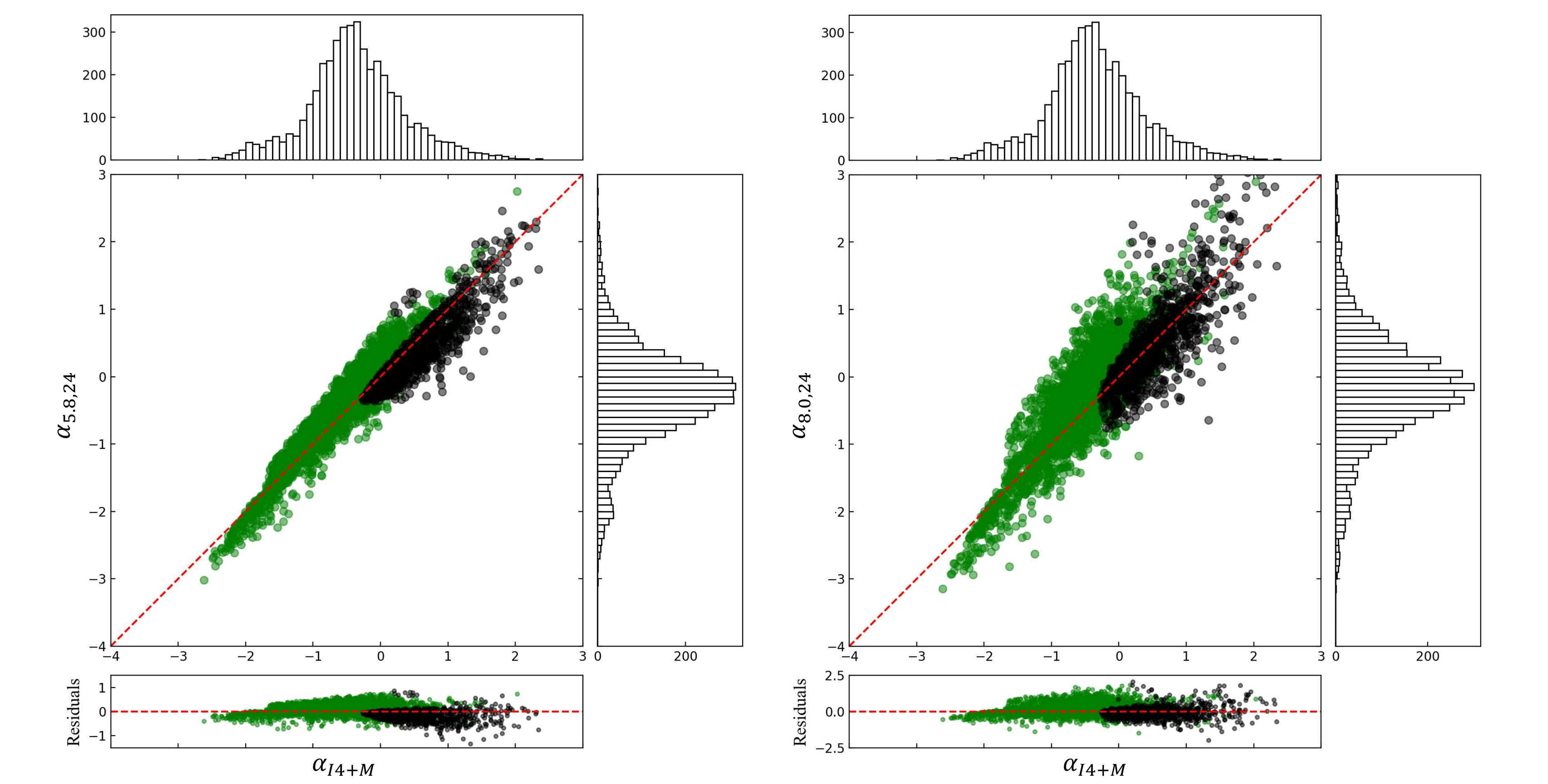}
\caption{correlations of SED slopes derived from two bands (MIPS 24 + one of the IRAC bands) and all five bands (MIPS 24 + IRAC 4 bands). Protostars (class 0 and I) are plotted in black while class II sources are plotted in green. The red dashed line shows the 1-to-1 relation.}
\label{f:abias}
\end{figure*}

By incorporating \fc\ photometry, the 19.7~$\mu$m and 31.5~$\mu$m fluxes become available, providing a possible substitute for the 24~$\mu$m data. In order to test this substitution, we have checked some two-point SED slope correlations with or without \fc\ data. The SED slopes between 4.5~$\mu$m (the second IRAC band) and 24.0~$\mu$m (MIPS 24 band) are compared with those between 4.5~$\mu$m and \fc\ bands at 19.7~$\mu$m or 31.5~$\mu$m. We specify the adopted bandpasses for all derived SED slopes. For instance, the SED slope between 4.5~$\mu$m and 24.0~$\mu$m is written as $\alpha_{4.5,24}$. The by-waveband $\alpha$ comparisons are plotted in Fig \ref{f:alpha} together with a 1-to-1 line. As shown in the figure, both $\alpha_{4.5,19.7}$ and $\alpha_{4.5,31.5}$ are correlated with $\alpha_{4.5,24}$, while $\alpha_{4.5,31.5}$ exhibits less scatter. In summary, \fc\ 31.5~$\mu$m photometry appears to constrain the mid-IR SED as well as MIPS 24~$\mu$m where both data sources are present, supporting confident use of \fc\ 31.5~$\mu$m data in combination with \sp\ IRAC data when MIPS 24~$\mu$m is unavailable or low quality.

Although the need for $\sim$20-30~$\mu$m photometry is the most obvious and essential, the selection of near-IR to IRAC bandpasses could also make a difference. We have followed the modern standard of ignoring JHK bands for estimating $\alpha$ in order to limit reddening bias \citep{muench2007}.
However, missing one or more IRAC bands could also result in some uncertainties. To further identify this so-called $\alpha$ bias, the SED slopes derived from two bands (MIPS 24 + one of the IRAC bands) are compared with those derived from all five bands (MIPS 24 + IRAC 4 bands). Only sources with well-sampled SED (requiring all five bands) are included in this analysis. As shown in Fig \ref{f:abias}, IRAC Channel 1 or 2 alone could guarantee a fairly good estimation of SED slope (RMS scatter = 0.06 and 0.09), which is not the case for Channel 3 or 4 (RMS scatter = 0.29 and 0.61).

\subsubsection{Interpolation for Missing Fluxes}

\begin{figure}
\includegraphics[width=1\linewidth]{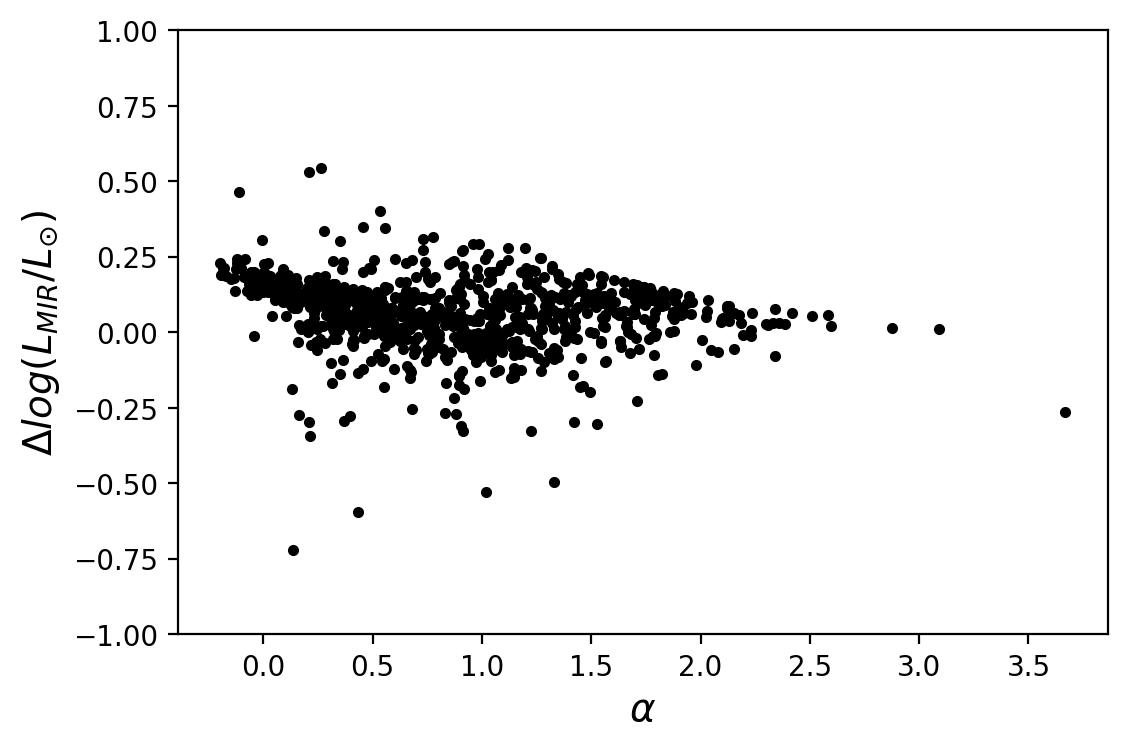}
\caption{The differences of $L_{\rm MIR}$ derived from real measurements or power-law interpolation. The sample includes all the SESNA protostars with one or more missing IRAC bands.}
\label{f:lmir_bias1}
\end{figure}

\begin{figure}
\includegraphics[width=1\linewidth]{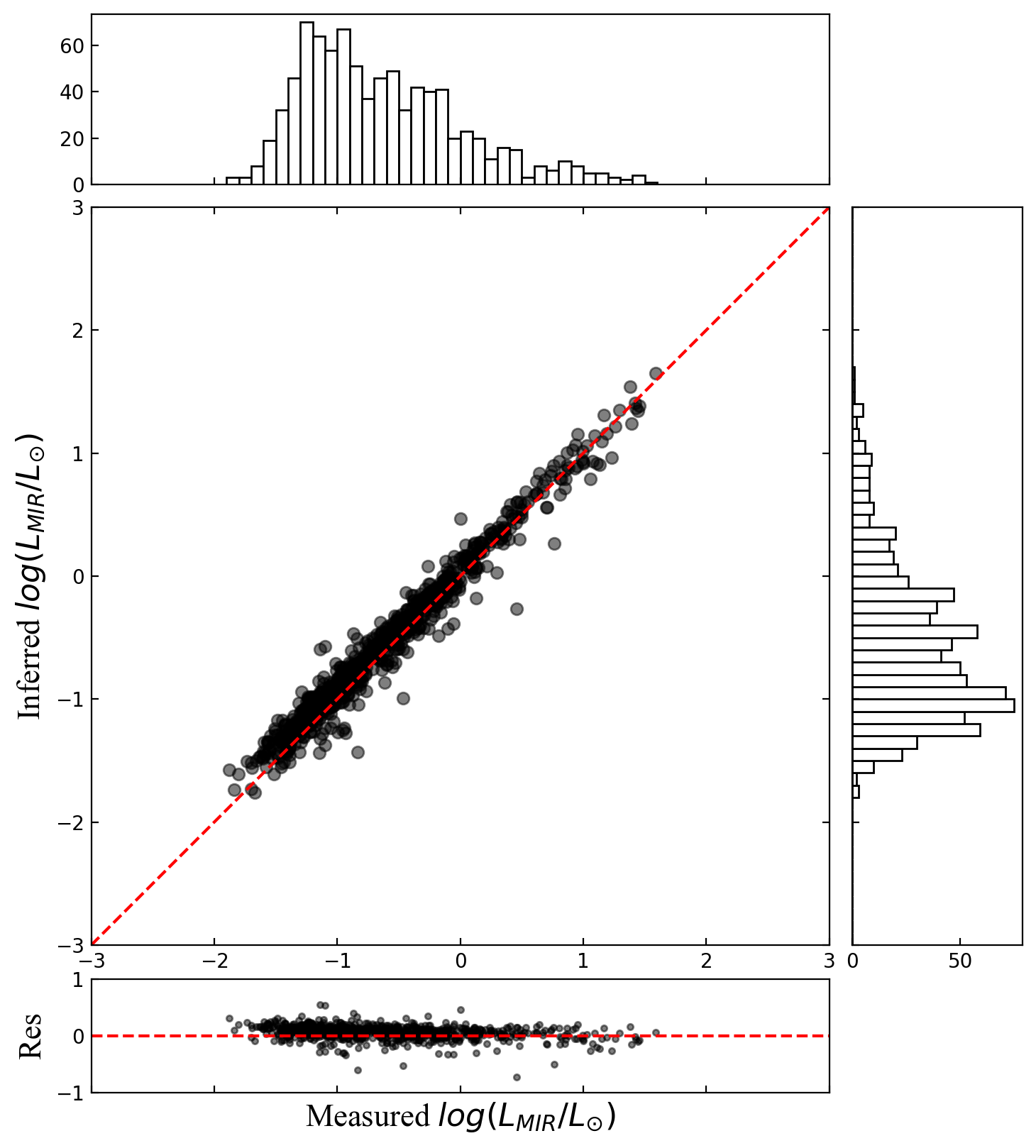}
\caption{Comparing $L_{\rm MIR}$ derived from real measurements or power-law interpolation for sources with well-sampled SED (requiring IRAC + M24).}
\label{f:lmir_bias2}
\end{figure}

Since $L_{\rm MIR}$ is determined by integrating fluxes from all bands, missing one or more bands in the integration could induce an underestimation of $L_{\rm MIR}$, and further affect the calculation of $L_{\rm bol}$. Making use of the MIPS 24 flux and a well-constrained SED slope, we are able to infer the fluxes at all the other bands by assuming a simple power-law SED shape and computing interpolated fluxes at each missing bandpass's wavelength. For SESNA protostars with one or more missing IRAC bands, we calculated $L_{\rm MIR}$ based on real measurements and the power-law interpolation respectively. The differences between these two results are plotted over $\alpha$ in Figure. \ref{f:lmir_bias1}, which indicates that dealing with those missing IRAC bands is important, especially for protostars with lower SED slopes.

To further test the validity of this interpolation method, we sorted sources with well-sampled SED (requiring all five bands) and compared the $L_{\rm MIR}$ derived from real measurements or inferred from $\alpha$. The good 1-to-1 correlation shown in Fig.\ref{f:lmir_bias2} indicates that the missing-band bias of $L_{\rm MIR}$ can be mitigated by inferring the missing band flux from $\alpha$. Since the J, H, and Ks bands contribute little to the integration of fluxes, requiring 5 bands (all 4 IRAC + MIPS 24 or \fc\ 31) could promise a relatively accurate calculation of $L_{\rm MIR}$. Note that in order to consistently use Eq. \ref{e:lmir}, we are interpolating the MIPS 24~$\mu$m photometry when we have \fc\ 31~$\mu$m data instead. As for our more inclusive sample that requires 2 bands, all the missing bands are interpolated and added into the integration, which results in an acceptable estimation of $L_{\rm MIR}$.

\subsubsection{Sample Selection}
The calculation of protostar luminosities largely relies on the calculation of SED slopes. In order to guarantee an accurate estimation of $\alpha$, all 4 IRAC bands plus the MIPS 24 band (or \fc\ 31 band as a substitution) are required in our 'clean' protostar sample for PLF calculation. By requiring the photometry in 5 bands and that the magnitude uncertainty of each band less than 0.2, we obtain a total of 981 protostars (953 from SESNA and 28 from \fc). We will refer to this protostar sample as Sample I afterwards.
Since IRAC Channel 1 or 2 alone with MIPS 24 band could guarantee a fairly good estimation of $\alpha$, we develop a more inclusive protostar sample by requiring only 2 bands, which will be referred as Sample II. The sample size is increased to 1212 protostars (1180 from SESNA and 32 from \fc).

When neither MIPS 24~$\mu$m nor \fc\ 31~$\mu$m data are available, the MIPS 24 90\% completeness flux, \fc\ 31~$\mu$m flux limit and WISE 22~$\mu$m flux are used for SED fitting, providing decent upper limits of $\alpha$ (see Section~ \ref{ss:comp} for details). Considering the double-beam contamination, the \fc\ Cycle~5 data had big uncertainties in source positions and photometry, thus were not used for deriving upper limits. In this way, we are missing less than 10 upper limits derived from \fc\ data, while limits from other approaches are still available. Focusing on all protostars in the \fc\ field of view of valid observations without MIPS 24~$\mu$m or \fc\ 31~$\mu$m detection, we are able to obtain their $1\sigma$ flux uncertainties at 31~$\mu$m through aperture photometry of the noise map (both the intrinsic error and the calibration error are taken into account). Then the $5\sigma$ flux uncertainties can be viewed as the flux upper limits for those protostars. An alternative way to calculate those limits is to simply use the \fc\ flux map, put apertures at the location of each undetected source and get the corresponding flux at 31~$\mu$m.

When 20-30~$\mu$m data are unavailable, it is also possible to constrain $\alpha$ with IRAC bands only. As shown in Fig \ref{f:alpha45}, $\alpha_{I4}$ seems to have a poor but non-negligible correlation with $\alpha_{I4+M}$, and thus might be usable for extrapolation to achieve a source of $L_{\rm bol}$ upper limits. With an offset of $\sim0.4$, $\alpha_{I4}$ would overestimate $\alpha_{I4+M}$ for 90\% of the sources, indicates that $\alpha_{I4} + 0.4$ is a reasonable upper limit of $\alpha$.

%%%%%%%%%%%%%%%%%%%%%%%%%%%%%%%%%%%%%%%%%%%%%%%%%%%%%%%%%%%%%%%%%%%%%%%%%%%%%%

\subsection{Completeness and Upper Limits} \label{ss:comp}

\begin{table*}
\caption{The upper limits of $\alpha$ and $L_{\rm bol}$ derived from different methods.}
\begin{filecontents*}{limit_begin.csv}
SESNA~Name,RA,DEC,\alpha_{FORCAST},Lbol_{FORCAST},\alpha_{COMPM24},Lbol_{COMPM24},\alpha_{WISE},Lbol_{WISE},\alpha_{IRAC},Lbol_{IRAC}
,(J2000),(J2000),,(L_{\odot}),,(L_{\odot}),,(L_{\odot}),,(L_{\odot})
J202540.03+365834.2,306.41679,36.976154,,,0.973,-0.398,2.584,0.349,1.373,0.116
J202539.75+365837.7,306.41562,36.977134,,,1.151,0.55,1.561,0.744,1.551,1.092
J202539.33+365844.4,306.41387,36.979011,,,-0.013,-0.26,1.579,0.492,0.387,-0.079
J202540.28+365845.1,306.41782,36.979192,,,0.061,-0.731,2.196,0.298,0.461,-0.48
J202540.20+365935.1,306.41749,36.99308,,,0.047,-0.127,1.087,0.391,0.447,0.109
J202537.30+365935.7,306.40542,36.993261,,,0.841,0.305,1.326,0.552,1.241,0.774
J202301.75+372230.4,305.75729,37.375099,,,0.062,-0.971,1.436,-0.285,0.462,-0.733
J202515.15+372245.6,306.3131,37.379344,,,0.603,-0.015,1.955,0.662,1.003,0.433
J202527.04+372327.7,306.36268,37.391041,,,0.604,0.99,1.342,1.389,1.389,1.57
J202526.20+372338.0,306.35916,37.393896,,,1.948,1.738,2.025,1.773,1.212,0.865
\end{filecontents*}
\csvreader[column count=11,
  no head,
  table head=\hline,
  late after last line=\\\hline,
  tabular=lllllllllll]{limit_begin.csv}
  {1=\one,2=\two,3=\three,4=\four,5=\five,6=\six,7=\seven,8=\eight,9=\nine,10=\ten,11=\eleven}
  {$\one$ & $\two$ & $\three$ & $\four$ & $\five$ & $\six$ & $\seven$ & $\eight$ & $\nine$ & $\ten$ & $\eleven$}
\label{tab:lim}

\vspace{0.1in}
Note: Limited to sources detected by SESNA. Only a portion of the full table is shown here to illustrate the form and content, the full table is available as supplementary material online.
\vspace{0.4in}
\end{table*}

Among a total of 2141 protostars, 927 of them have no MIPS 24 photometry, which means that a large fraction of SESNA protostars cannot be used to derive $\alpha$ and $L_{\rm bol}$ because of lack of data. Nevertheless, we can get luminosity upper limits for these protostars in a variety of ways, and in most cases, these limits will enable us to push these sources down to well-populated portions of the PLF where their presence or absence will not strongly affect the shape of the function. 

Any flux upper limit in the 20-30~$\mu$m range is able to put an upper limit on $\alpha$, and thus has potential as a strong luminosity constraint. The 90\% MIPS 24 completeness provides the first approach to lay an upper limit on both $\alpha$ and $L_{\rm bol}$. The second way is using the \fc\ 31~$\mu$m fluxes and uncertainties. As discussed in Section~ \ref{ss:yso}, both the $5\sigma$ flux uncertainties and the corresponding fluxes can be viewed as the flux upper limits at 31~$\mu$m. The upper limits of $\alpha$ and $L_{\rm bol}$ can then be derived assuming a power-law SED shape. The higher resulting upper limit of these two methods can be safely determined as the $L_{\rm bol}$ limit derived from \fc\ data. The third approach is to make use of the WISE data, where the MIPS 24~$\mu$m flux is replaced by the WISE 22~$\mu$m flux in calculation. Another alternative way is to derive the upper limit of $\alpha$ from adjusting the $\alpha_{I4}$ and then use the results to calculate $L_{\rm bol}$ (see Section \ref{ss:yso}). 

\begin{figure}
\includegraphics[width=1\linewidth]{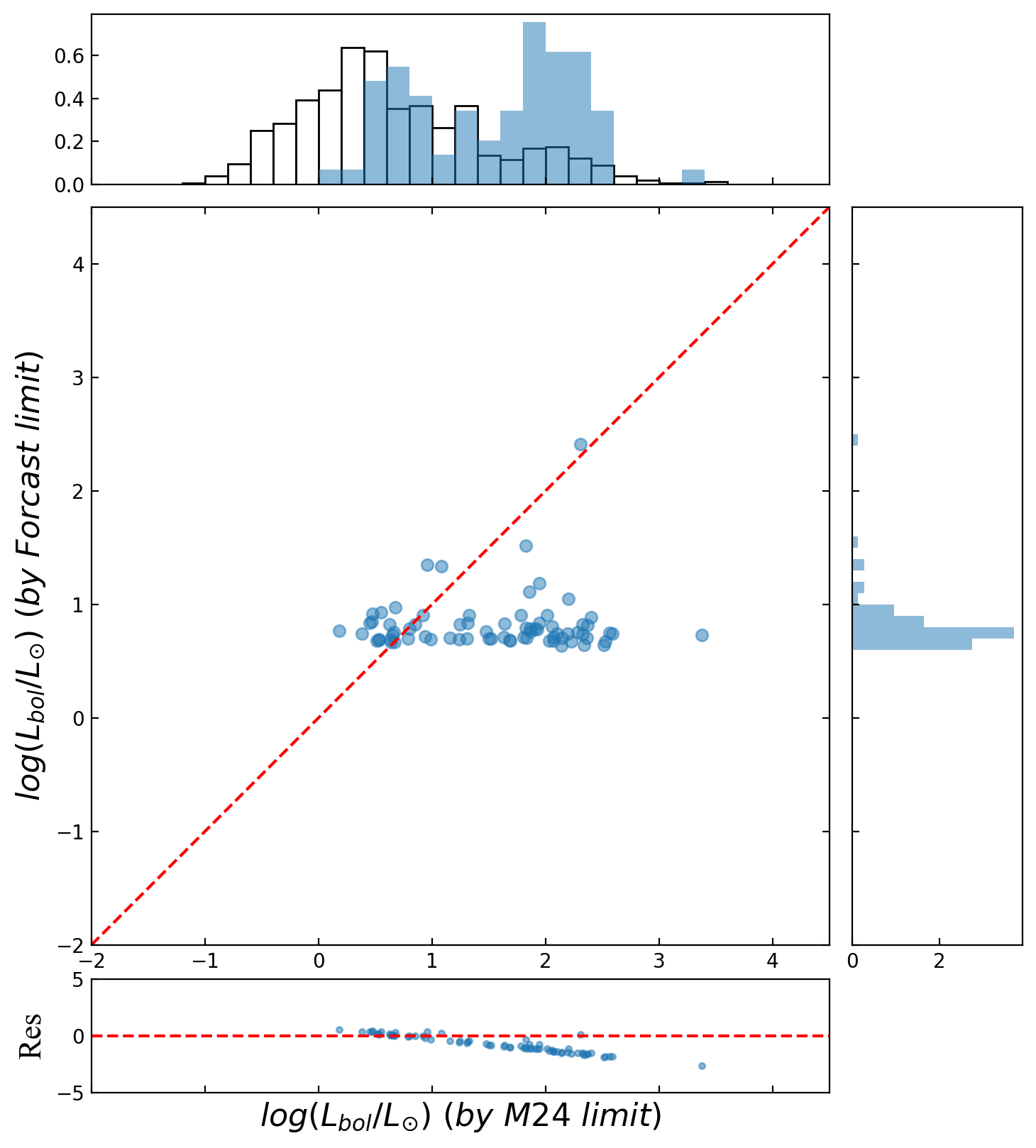}
\caption{The upper limits of protostar luminosity derived from MIPS 24 completeness or \fc\ limits. The red dashed line shows the 1-to-1 relation and the bottom panel shows the differences of the two axis. The distribution histogram of each upper limits is plotted on the corresponding axis, where the white bins include all the protostars with MIPS 24 completeness data, and the blue bins further require the \fc\ coverage.}
\label{f:limfc}
\end{figure}

\begin{figure}
\includegraphics[width=0.9\linewidth]{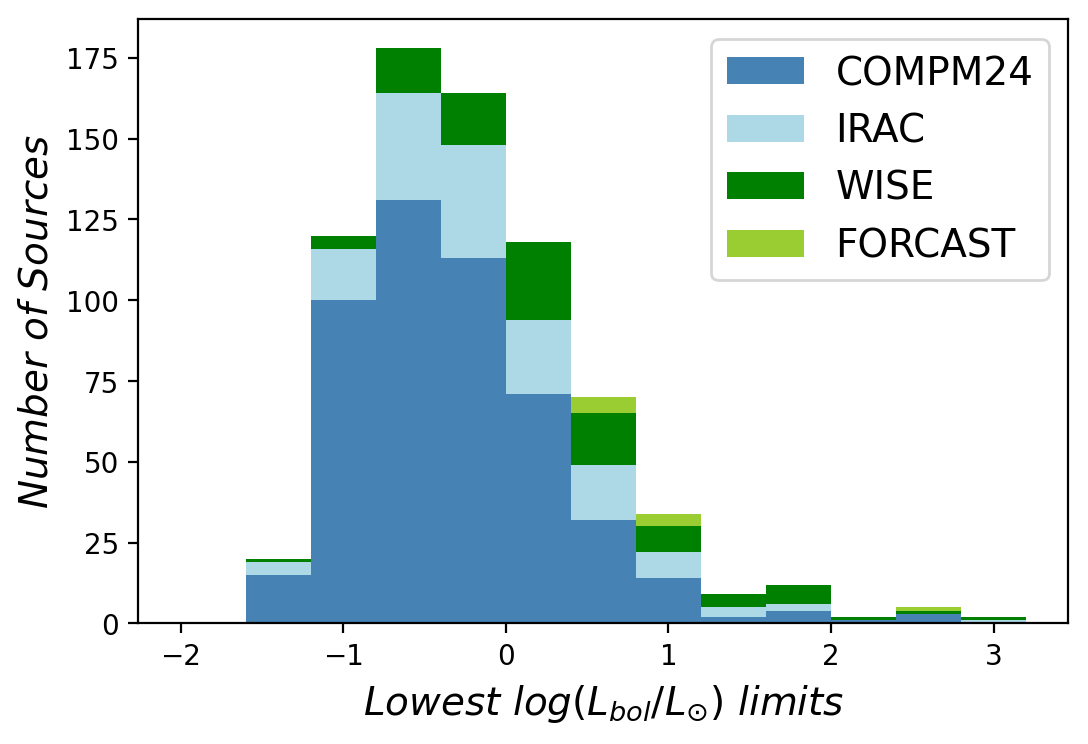}
\caption{Stacked histogram of best $L_{\rm bol}$ upper limits.}
\label{f:lim}
\end{figure}

We present a comparison of $L_{\rm bol}$ limits derived from MIPS24 completeness or \fc\ data in Fig \ref{f:limfc}. A few sources fall on the 1-to-1 line, while most of the sources sit on the right and there is an obvious threshold on the $y$ axis. Since the \fc\ beam is much cleaner than the MIPS beam at large radius, its sensitivity gain among bright sources and nebulosity is frequently superior to MIPS. We have leveraged this benefit through field selection and managed to push the $L_{\rm bol}$ upper limits down to $log(L_{\rm bol}/L_{\odot})\sim0.5$.

By now, we have introduced four different ways to provide upper limits on the $\alpha$ and $L_{\rm bol}$ of selected sources, the results are compared in Table \ref{tab:lim}. When multiple approaches are available for a single source, the lowest (most constrained) upper limit of $L_{\rm bol}$ is taken as the final limit. Fig \ref{f:lim} shows the distribution of the best upper limits and their respective approaches. In most circumstances, the MIPS 24 completeness provides the best constraint of $L_{\rm bol}$, but \fc\ limits play a significant role in constraining the luminosity of sources in bright or confused regions. In summary, we obtain 735 valid upper limits for both $\alpha$ and $L_{\rm bol}$, which is $\sim$75\% of the size of Sample I.

%%%%%%%%%%%%%%%%%%%%%%%%%%%%%%%%%%%%%%%%%%%%%%%%%%%%%%%%%%%%%%%%%%%%%%%%%%%%%%

\subsection{The Protostellar Luminosity Function}\label{ss:plf}

The Protostellar Luminosity Function (PLF) is simply the probability density distribution of the luminosities of all detected protostars per logarithmic bin of Lbol ($\psi_p(L)=$ dN/dlog$L_{\rm bol}$). We derive errors by assuming that Poisson counting statistics hold for the number of protostars in each bin. In this work, we aim to better characterize the ratio of the number of high ($1.5<{\rm log}~L_{\rm bol}<2.5$) to intermediate ($0.5<{\rm log}~L_{\rm bol}<1.5$) luminosity protostars across a variety of measurably different natal environments. To do so, we fit a simple power law ($y=ax^b$ where $y=$dN/dlog$L_{\rm bol}$ and $x=$log$L_{\rm bol}$) within this luminosity range and compare the power law index ( $b$ values) in different cases. The power law function form is also supported by theoretical models (see Section \ref{ss:model}). To get the best fitting results as well as the standard deviations, we use the nonlinear least square optimization technique in the Python 'curve\_fit' function in the SciPy module.

\begin{figure}
\includegraphics[width=1\linewidth]{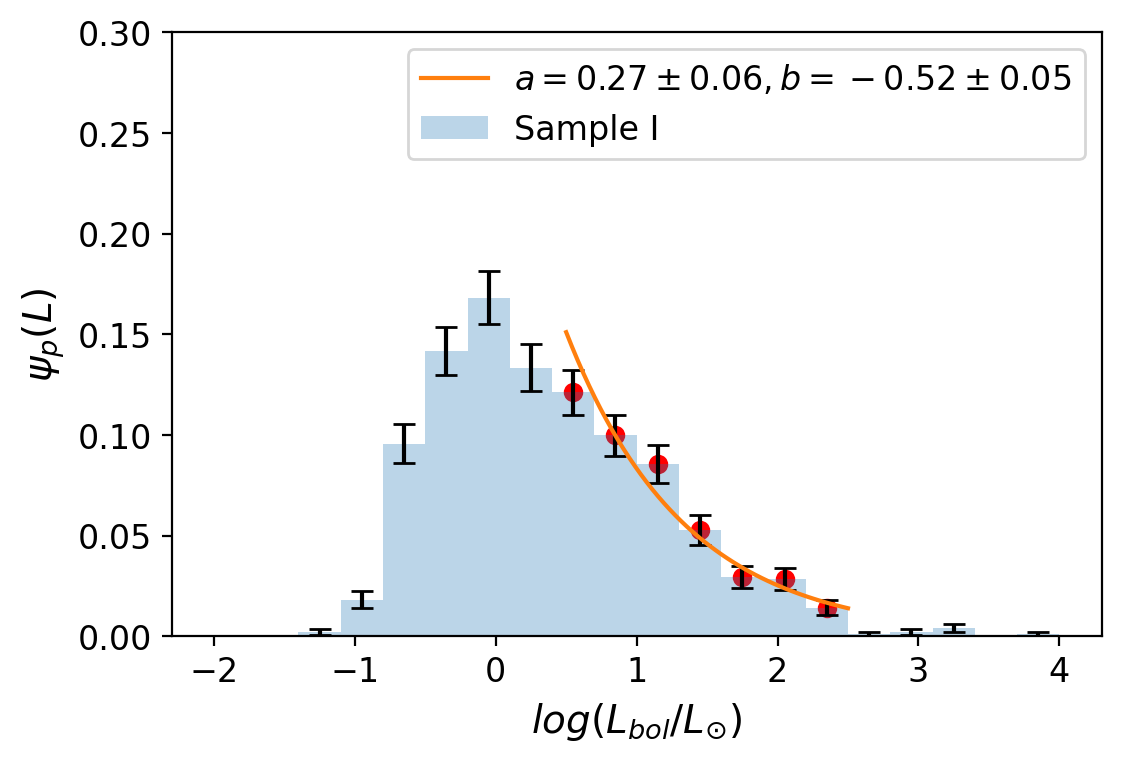}
\includegraphics[width=1\linewidth]{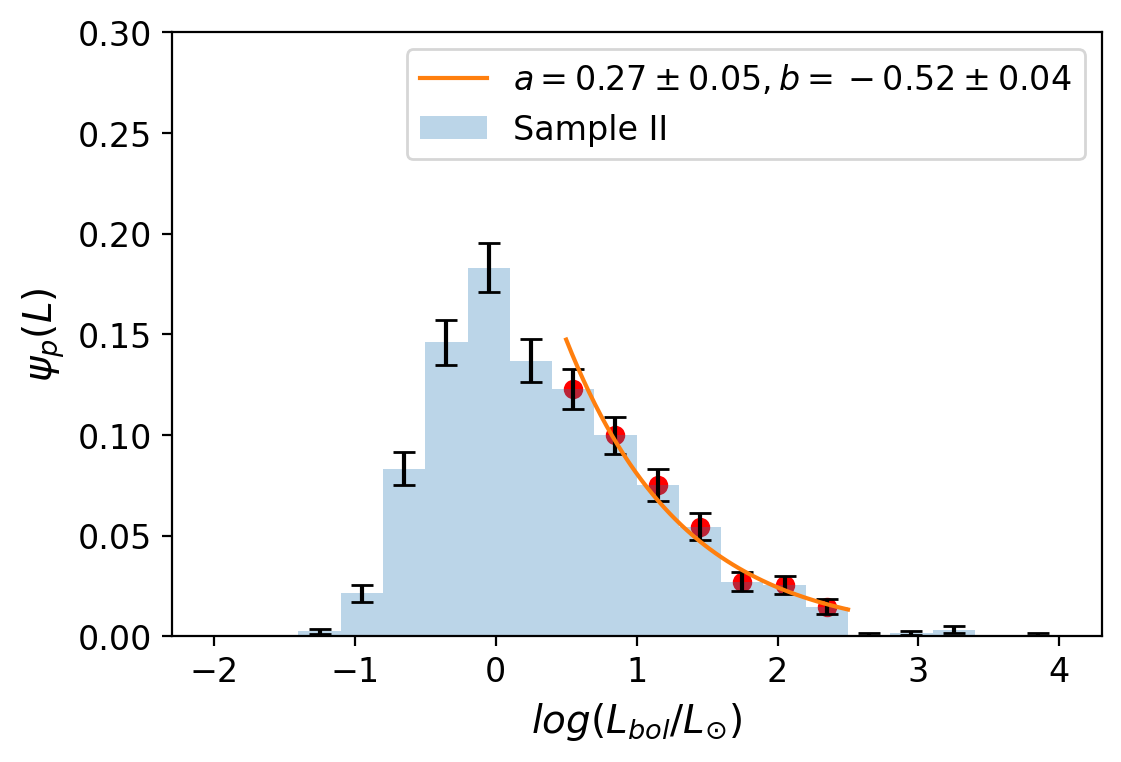}
\caption{PLFs for Sample I and Sample II along with the best power-law fits.}
\label{f:plf}
\end{figure}

\begin{figure}
\includegraphics[width=1\linewidth]{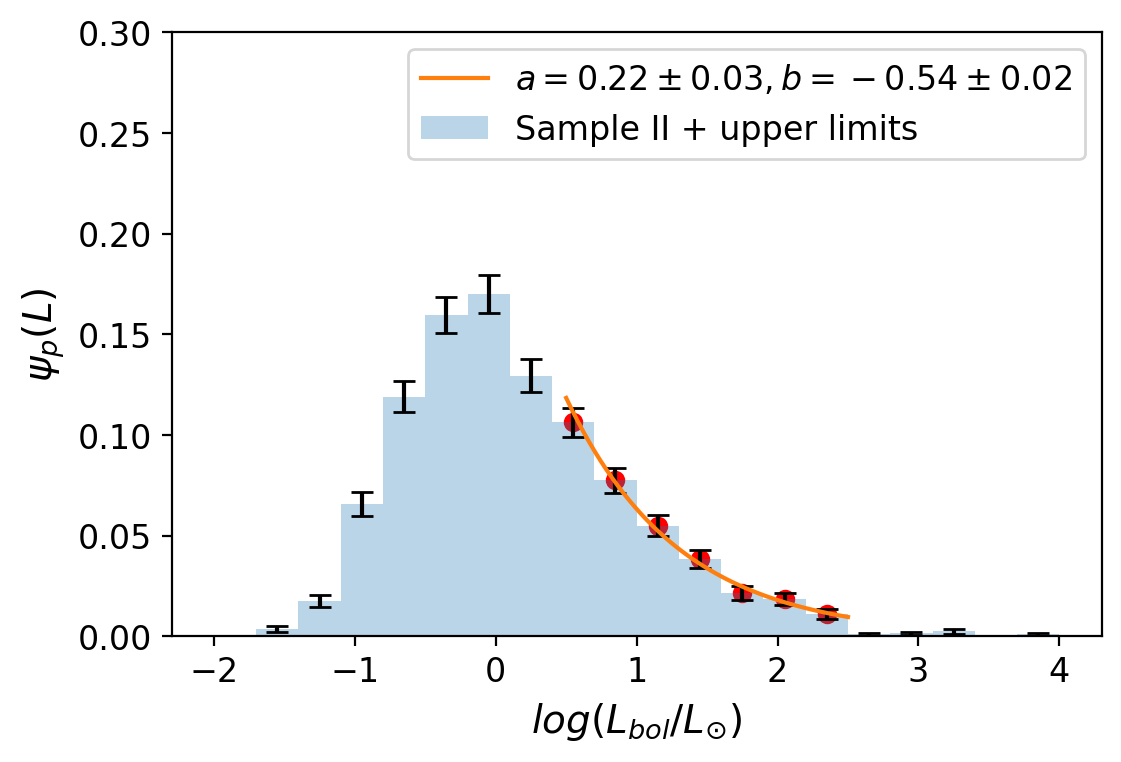}
\caption{PLF after adding valid luminosity upper limits to Sample II along with the best power-law fits.}
\label{f:plf_lim}
\end{figure}

\begin{figure}
\includegraphics[width=1\linewidth,trim={0 0 20 0}]{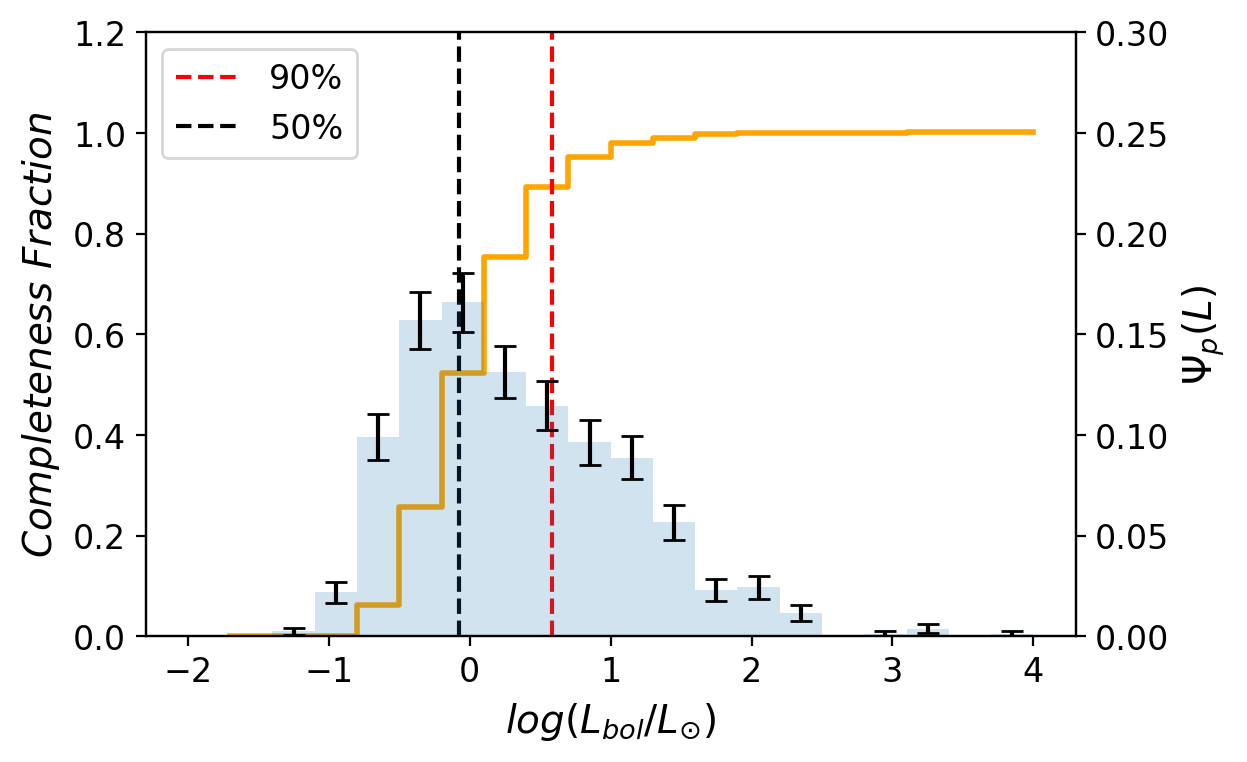}
\includegraphics[width=1\linewidth,trim={0 0 20 0}]{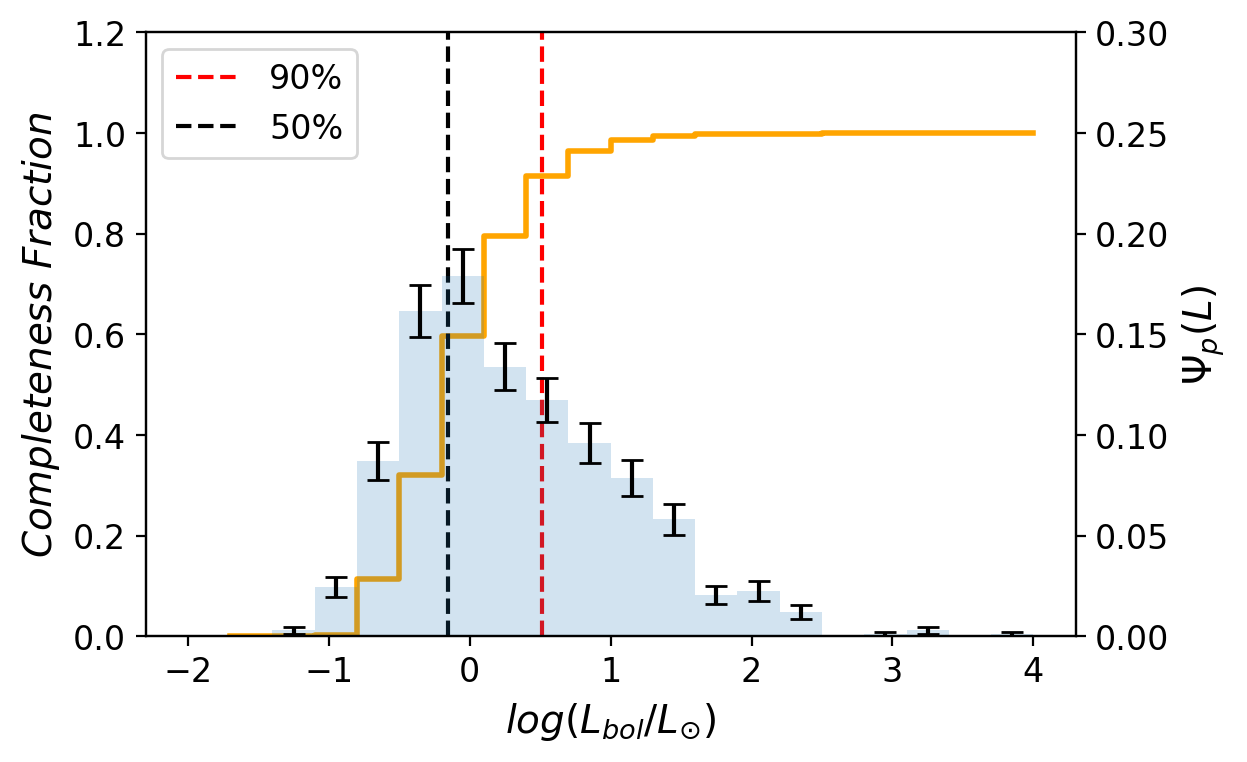}
\caption{Completeness fraction decay curves for Sample I and Sample II overlaid on corresponding PLFs. The black and red dashed lines mark the 50\% and 90\% completeness luminosities respectively.}
\label{f:frac}
\end{figure}

According to the power-law fitting within the range of $0.5<log(L_{\rm bol}/L_{\odot})<2.5$, the power law index $b$ is determined as $-0.52\pm0.05$ for Sample I (Fig \ref{f:plf}). By requiring only 2 bands, Sample II results in a similar shape of PLF with $b=-0.52\pm0.04$. For comparison, we also fit to the PLF from a sample of 2007 protostar candidates given by \citealt{kryu2014}. Although the sample selections have substantial differences, the overall shapes of the PLF for $log(L_{\rm bol}/L_{\odot})>0.5$ are comparable. Considering the uncertainties, the fitting results are consistent with each other (Table \ref{tab:3range}).

\subsubsection{Potential Incompleteness}
As noted above, many protostars go undetected at $\sim$20-30~$\mu$m, but we can still place reliable upper limits on those fluxes and by extension, their luminosities. Given that the distribution of luminosity upper limits is skewed to rather low values (typically $L<10L_{\odot}$), any sources with luminosities that are substantially dimmer than their upper limits would exit the PLF fitting range we have adopted. Thus, to gauge the maximal impact that omitting these sources from our PLF analyses might have, we simply add the upper limit luminosities to Sample II and perform the same power law fit.
After enclosing 735 valid upper limits, we are able to extend the sample size from 1212 to 1947. As shown in Fig \ref{f:plf_lim}, the best-fit power-law slope becomes $-0.54\pm0.02$, which is highly consistent with the result of Sample II (less than $1\sigma$ difference) with a significantly smaller uncertainty.

\begin{figure*}
\includegraphics[width=1\linewidth,trim={10 30 10 10}]{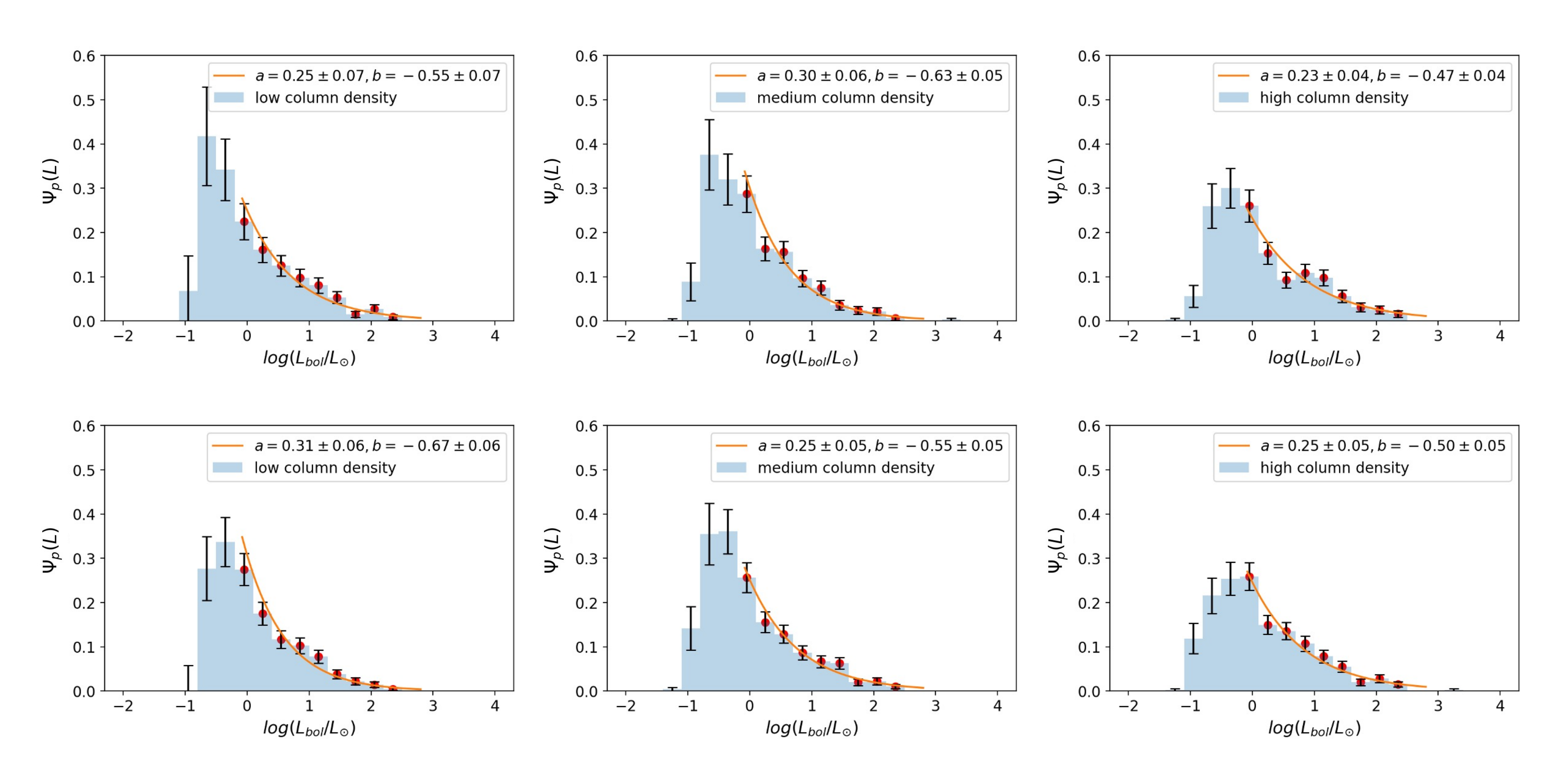}
\caption{The PLFs of different gas column density ranges after correcting for incompleteness. The first row shows the 981 sources from Sample I, while the second row shows 1212 sources from Sample II.}
\label{f:comp_corr}
\end{figure*}

Next we explore a more comprehensive way to account for the protostellar incompleteness, where we use the SESNA by-bandpass point source completeness products to derive the luminosity from $\alpha$. The SESNA completeness map provides spatially resolved 90\% differential completeness limit in magnitudes sliced by bandpasses. Through a full sampling of the $L_{\rm MIR}$ bands required with the power law SED while stepping through the $\alpha$ space, we are able to obtain the luminosity cube. Given the location of each protostar and its $\alpha$ value, we will be able to extract its completeness luminosity and then combine the results to build a completeness fraction decay curve (Fig \ref{f:frac}). The bin size is consistent with our PLF, and the 50\% and 90\% completeness luminosities come from an interpolation of the step function. At lower luminosity, the detection completeness is relatively low, which means that part of the protostars are not included due to the detection limit. To test the effect of these missing sources, we conducted the completeness correction by scaling up each PLF bin according to its completeness fraction, and then compared with the original results. The completeness cubes for Sample I and Sample II were generated separately. However, after completeness correction, the PLFs of these two different samples are highly consistent. After correction, the luminosity range for power-law fitting can be appropriately extended to the lower luminosity end (e.g. $L_{\rm bol}=1~L_{\odot}$), but the best-fit power-law slope doesn't change much ($b=-0.48\pm0.03$ for Sample I and $b=-0.49\pm0.03$ for Sample II). We also conduct the completeness correction for different subsets of PLFs, the results are shown in Table \ref{tab:3range}. As an example, the corrected PLFs with different gas column density ranges for both samples are shown in Fig \ref{f:comp_corr}. Interestingly, the variation in best-fit power-law slopes for different gas column density ranges becomes smaller after the correction, while the stellar-density-based subdivision still shows a prominent difference among different ranges. This discrepancy may arise from the complexity in the star-gas correlation, such as projection effects of the gas column density.

Another potential incompleteness arises from the source classification. As shown in Fig \ref{f:a-lbol}, Class II sources with $\alpha\ge-0.3$ and unclassified sources with $log(L_{\rm bol})\ge-1$ overlap well with protostars. Although the SESNA classification is optimized, there might be some missing protostars hidden among the Class IIs and unclassified SESNA sources in the analysis above. We believe that the majority of the Class IIs with nearly Flat spectrum should be edge-on disks, while a small fraction of them may actually be protostars. To address the maximal impact the misclassification could have on the PLF shape, we explored the worst case scenarios by including all these potential missing (misclassified) sources in our original protostar catalogue. In this way, the sample size will increase by $\sim 30\%$, though the PLF power-law slopes show very little change. Thus, the source classification shows a very limited impact on the completeness of our sample.
%%%%%%%%%%%%%%%%%%%%%%%%%%%%%%%%%%%%%%%%%%%%%%%%%%%%%%%%%%%%%%%%%%%%%%%%%%%

\subsubsection{Possible Contamination}
\label{ss:contam}

\begin{figure}
\includegraphics[width=1\linewidth]{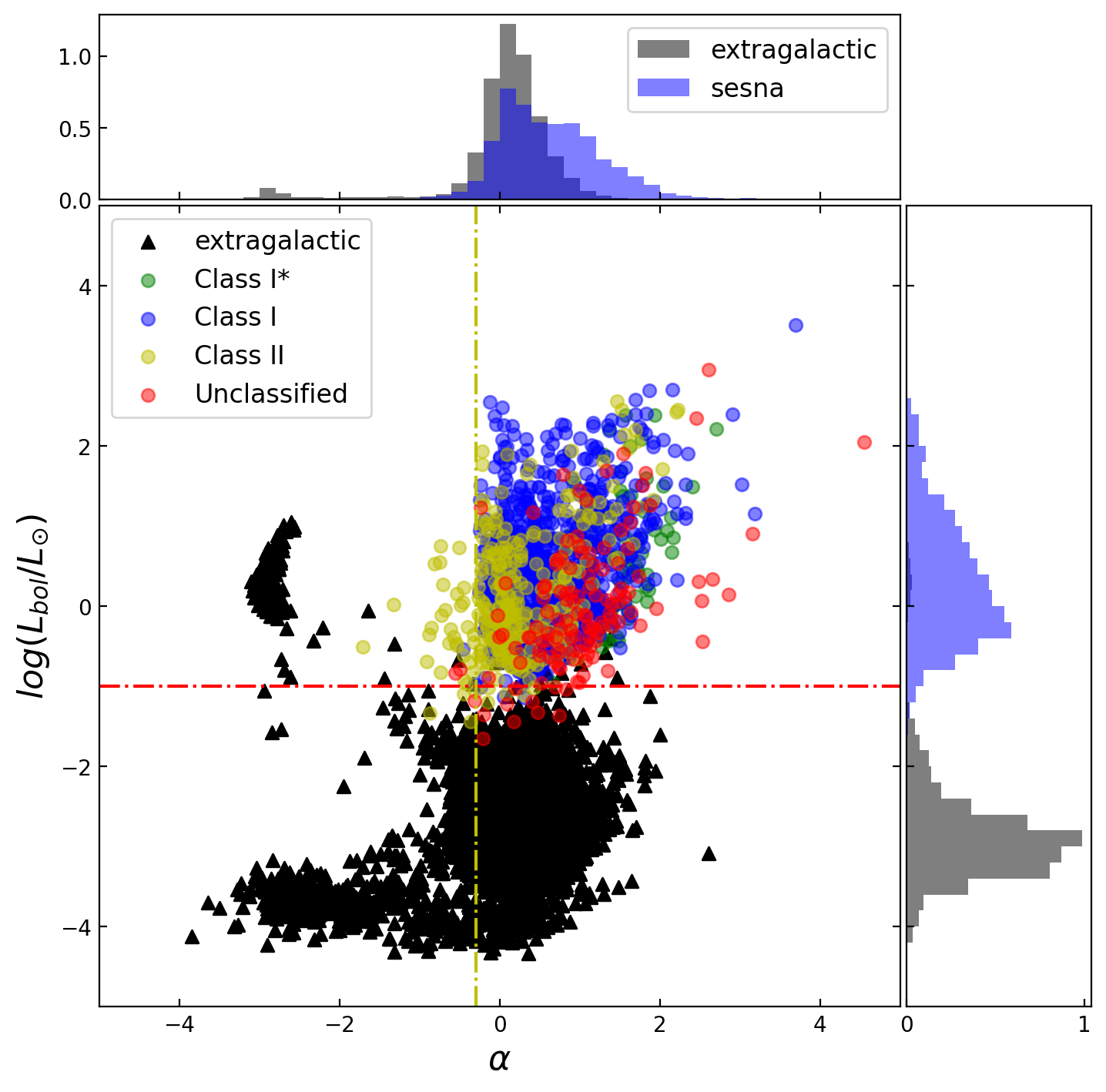}
\caption{Distribution of different classes of SESNA YSOs and extragalactic sources in the $\alpha-L_{\rm bol}$ space. The vertical line shows the threshold at $\alpha=-0.3$, and the horizontal line marks $log(L_{\rm bol})=-1$.}
\label{f:a-lbol}
\end{figure}

Contaminants are a known issue for YSO census work (e.g. \citealt{guter2009, megeath2012, pokhrel2020}), and thus their potential impact on the PLF characterizations should be considered. Possible contamination to the PLF includes edge-on disks (Class IIs), galaxies and reddened Class II objects. As shown in Fig \ref{f:a-lbol}, the SESNA YSOs and extra-galactic sources (including some field stars) show good separation on the $\alpha-L_{\rm bol}$ plot, which indicates that the galaxy contamination is not a big concern.

First, we look into the edge-on disk (Class II) contamination. An edge-on YSO which has dissipated its envelope (Stage II) but is viewed through its disk may have reddened IR emission and a flatter SED, which results in an $\alpha$ typical of a Class I source \citep{robi2006}. By requiring detections in the MIPS 24 band, we have ruled out most Class II sources, but edge-on disk contamination might still play a role in low extinction regions, or anywhere with a high concentration of Class II YSOs. To take this contamination into account, we need to remove Class II sources with Flat spectrum in our PLF. The adopted definition of Flat spectrum is $-0.3 < \alpha < 0.3$ given by \citealt{greene1994}. As shown in Appendix. \ref{a:1}, the upper limit of the edge-on disk contamination rate is estimated as $\sim$3.9\%. Here we adopt this upper limit to test the maximum effect of the edge-on disk contamination. Given this fraction, within a total of 3044 Class II sources with valid 20-30~$\mu$m photometry identified by SESNA in \cygx, $\sim$76 are misclassified as Class I and included in our protostar sample, giving a protostellar contamination rate of $\sim$7.7\%. Over half of the Class II contamination lies in the low column density range, it can hardly make any change to the PLF shape at medium or high column densities. By removing the estimated contamination from Sample I, the best-fit power-law slope becomes $-0.45\pm0.06$, which is less than $1\sigma$ difference from the original fit. We went on to remove estimated Class II contamination for different sub-samples of PLF. The effect of Class II contamination on PLFs in different environments is also minor, which only changes the best-fit power-law slope by $\sim$0.01.

Another issue is the reddened Class II contamination. Given the SESNA source classification process, the contamination from reddened Class IIs is small and skewed toward low luminosities. As demonstrated in Appendix. \ref{a:2}, the reddened Class II contaminants in the low column density sample are negligible, the protostellar contamination rate is $\sim$2.1\% for medium column density and $\sim$3.4\% for high column density. This rate is much smaller than the edge-on disk contamination rate, and can hardly make any change to the PLF fits.

%%%%%%%%%%%%%%%%%%%%%%%%%%%%%%%%%%%%%%%%%%%%%%%%%%%%%%%%%%%%%%%%%%%%%%%%%%%

\subsubsection{PLF and the Local Environment}
\label{ss:env}

\begin{figure*}
\includegraphics[width=1\linewidth,trim={10 30 10 10}]{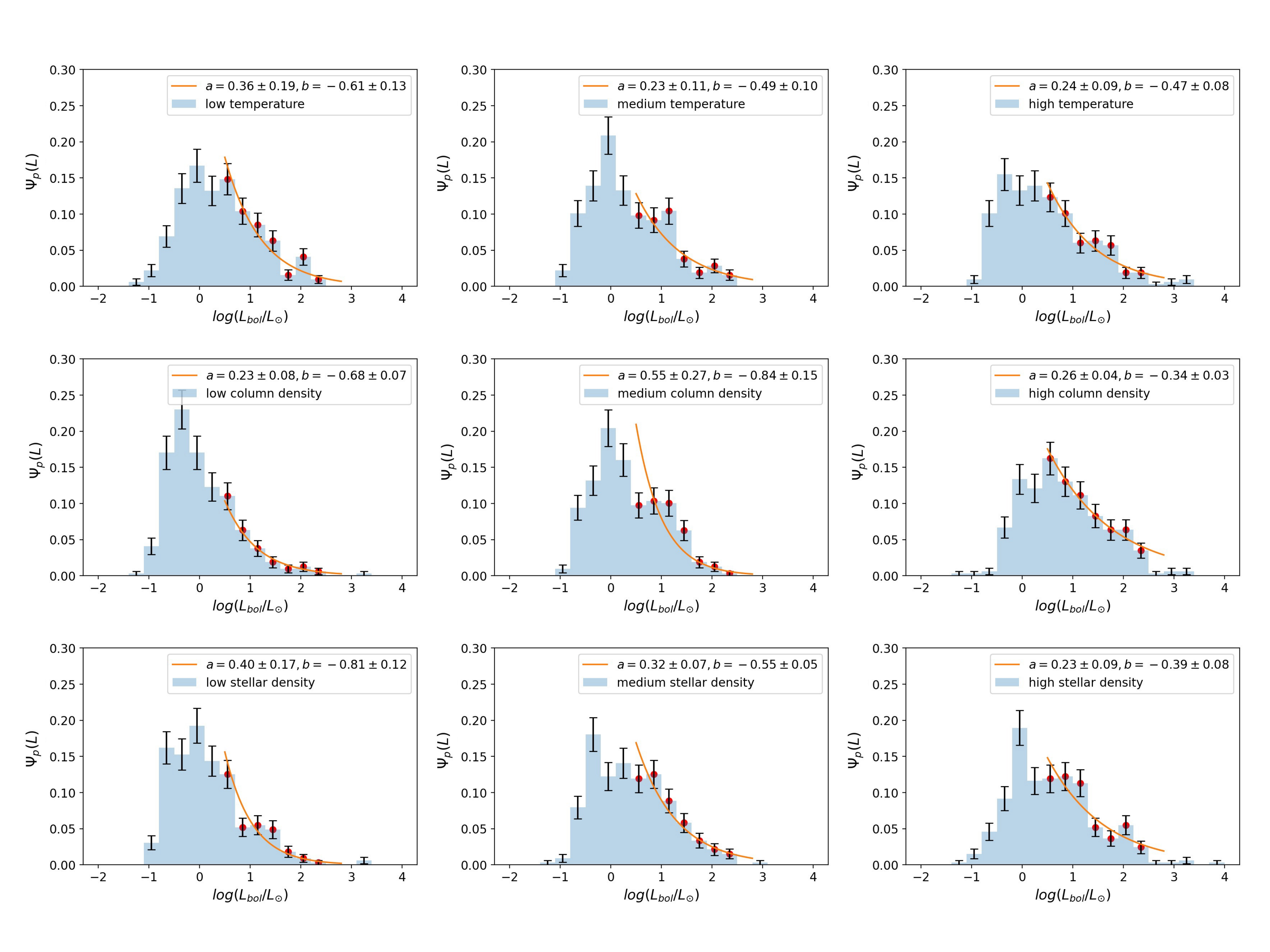}
\caption{PLFs for evenly split sub-samples of Sample I according to different gas temperature, gas column density and stellar density. The best-fit power-law curves and parameters are included for each case.}
\label{f:3range}
\end{figure*}

Next we split the original sample into several sub-samples to study how the PLF may be affected by the local environment, including gas temperature, gas column density and stellar density. We evenly split Sample I into three sub-samples of 327 sources according to each physical property, the parameter ranges for all sub-samples are listed in Table \ref{tab:3range}. The temperature and column density comes from Herschel observations described in Section \ref{ss:arx}, and the stellar density (local surface density) is characterized by the fourth nearest-neighbor distance \citep{kryu2012}. Note that we derived the distances within our pure protostar samples, while \citealt{kryu2012} and \citealt{kryu2014} used the entire YSO sample, including all Class II and protostar candidates. For each physical property, we divide the whole sample into three sub-samples of the same size and conduct power-law fittings. As discussed in previous sections, the SED slope has a profound effect on the calculation of $L_{\rm bol}$. To make sure that the sample division is not strongly biased to $\alpha$, we checked the ratio of low $\alpha$ sources to high $\alpha$ sources (divided by the median value) for each sub-sample. The ratio varies from 0.8 to 1.2, and there is no big discrepancy among different sub-samples. Then we compared the best-fit power-law slopes between different sub-samples and with the total sample. Table \ref{tab:3range} summarizes the results for different samples and different physical property ranges, including Sample I and Sample II with and without completeness correction as well as the sample taken from \citealt{kryu2014}. The threshold refers to the boundary of each range, and $b$ represents the best-fit power-law slope. The differences between the $b$ values of the lower and higher ranges are listed in the last column.

\begin{table*}
\caption{Best-fit power-law slopes of the PLFs for each range of physical properties and each sample.}
\scriptsize
\begin{tabular}{c|l|llllllll}
\hline
\multicolumn{1}{l|}{} & Sample & Low & Medium & High & $b_{tot}$  & $b_{low}$ & $b_{med}$ & $b_{high}$ & $\Delta\ b_{high-low}$ \\ \hline
\multirow{3}{*}{\begin{tabular}[c]{@{}c@{}}Temperature\\ T(K)\end{tabular}} & Sample I (981) & 15.7-21.3 & 21.3-23.8 & 23.8-37.5 & $-0.52\pm0.05$ & $-0.61\pm0.13$ & $-0.49\pm0.10$ & $-0.47\pm0.08$ & $0.14\pm0.15$  \\
 & Sample I (corrected) & & & & $-0.48\pm0.03$ & $-0.48\pm0.04$ & $-0.57\pm0.04$ & $-0.64\pm0.10$ & $-0.16\pm0.11$  \\
 & Sample II (1212) & 15.7-21.3 & 21.3-23.8 & 23.8-37.5 & $-0.52\pm0.04$ & $-0.67\pm0.15$ & $-0.53\pm0.06$ & $-0.43\pm0.05$ & $0.24\pm0.16$  \\
 & Sample II (corrected) & & & & $-0.49\pm0.03$ & $-0.54\pm0.06$ & $-0.56\pm0.04$ & $-0.60\pm0.05$ & $-0.06\pm0.08$  \\
 & Kryukova (2007) & 15.7-22.3 & 22.3-25.3 & 25.3-38.9 & $-0.53\pm0.03$ & $-0.65\pm0.05$ & $-0.53\pm0.08$ & $-0.43\pm0.02$ & $0.22\pm0.05$  \\ \hline
\multirow{3}{*}{\begin{tabular}[c]{@{}c@{}}Column Density\\ $lgN(cm^{-2})$\end{tabular}} & Sample I (981) & 20.5-22.0 & 22.0-22.2 & 22.2-23.6 & $-0.52\pm0.05$ & $-0.68\pm0.07$ & $-0.84\pm0.15$ & $-0.34\pm0.03$ & $0.34\pm0.08$  \\
 & Sample I (corrected) & & & & $-0.48\pm0.03$ & $-0.55\pm0.07$ & $-0.63\pm0.05$ & $-0.47\pm0.04$ & $0.08\pm0.08$  \\
 & Sample II (1212) & 20.5-22.0 & 22.0-22.2 & 22.2-23.6 & $-0.52\pm0.04$ & $-0.73\pm0.05$ & $-0.93\pm0.14$ & $-0.33\pm0.02$ & $0.40\pm0.05$  \\
 & Sample II (corrected) & & & & $-0.49\pm0.03$ & $-0.67\pm0.06$ & $-0.55\pm0.05$ & $-0.50\pm0.05$ & $0.17\pm0.08$  \\
 & Kryukova (2007) & 21.2-22.0 & 22.0-22.2 & 22.2-23.9 & $-0.53\pm0.03$ & $-0.70\pm0.07$ & $-0.77\pm0.09$ & $-0.39\pm0.03$ & $0.31\pm0.08$  \\ \hline
\multirow{3}{*}{\begin{tabular}[c]{@{}c@{}}Stellar Density\\ Dnn4(pc)\end{tabular}} & Sample I (981) & 0.2-2.3 & 2.3-4.3 & 4.3-20.8 & $-0.52\pm0.05$ & $-0.81\pm0.12$ & $-0.55\pm0.05$ & $-0.39\pm0.08$ & $0.42\pm0.14$  \\
 & Sample I (corrected) & & & & $-0.48\pm0.03$ & $-0.74\pm0.08$ & $-0.47\pm0.04$ & $-0.39\pm0.05$ & $0.35\pm0.09$  \\
 & Sample II (1212) & 0.2-1.9 & 1.9-4.0 & 4.0-21.0 & $-0.52\pm0.04$ & $-0.80\pm0.13$ & $-0.57\pm0.06$ & $-0.39\pm0.05$ & $0.41\pm0.14$  \\
 & Sample II (corrected) & & & & $-0.49\pm0.03$ & $-0.74\pm0.09$ & $-0.48\pm0.05$ & $-0.41\pm0.03$ & $0.33\pm0.09$  \\
 & Kryukova (2007) & 0.1-1.3 & 1.3-2.5 & 2.5-21.3 & $-0.53\pm0.03$ & $-0.75\pm0.05$ & $-0.55\pm0.03$ & $-0.44\pm0.06$ & $0.31\pm0.08$  \\ \hline
\end{tabular}
\label{tab:3range}
\end{table*}

\begin{figure*}
\includegraphics[width=1\linewidth,trim={0 40 0 10}]{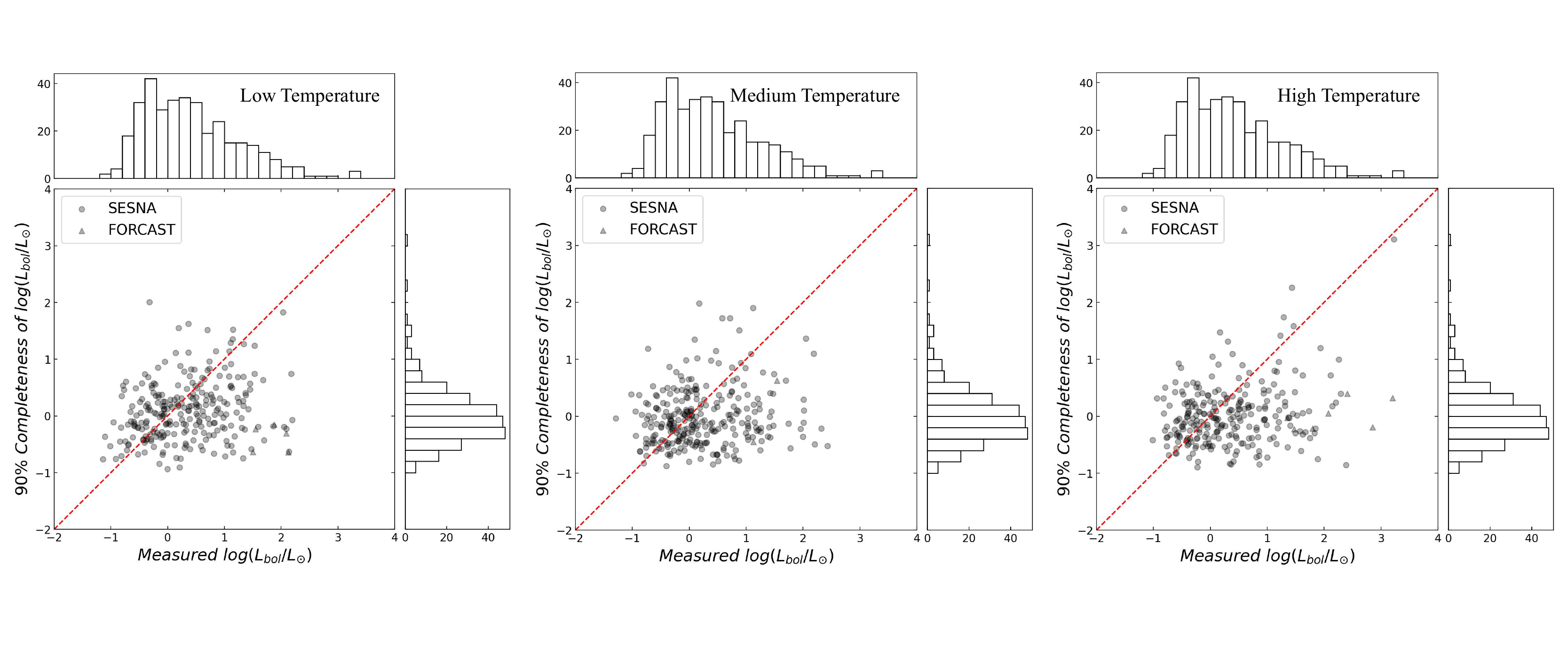}
\caption{The 90\% completeness $L_{\rm bol}$ over the real measured $L_{\rm bol}$ for different gas temperature ranges. The division on temperature is identical to that used for generating PLFs in Fig \ref{f:3range}. The histograms of $L_{\rm bol}$ and completeness $L_{\rm bol}$ are shown on the $x$ and $y$ axis, and a 1-to-1 line is plotted in red. Sources from the SESNA catalogue are marked as circles, while sources observed by \fc\ only are marked as triangles.}
\label{f:comp_t}
\end{figure*}

Within our fitting range, the PLF shows little variation with local dust temperature. The difference between the best-fit power-law slopes is less than $1\sigma$ for low and high temperature range. Meanwhile, the gas column density and stellar density have some effect on the PLF. The best-fit power-law slope tends to be larger at high gas column density or high stellar density, which means the PLF tail tends to be flatter. The slope differences between the low and high density sub-sample are close to $3\sigma$ in both cases. We conclude that the PLFs in regions of either high gas column density or high stellar density have more high-luminosity sources.

It is possible that the effects of incompleteness are different for different ranges of physical properties. To explore this issue in more detail, we show an example plot of 90\% completeness $L_{\rm bol}$ over the real measured $L_{\rm bol}$ for our samples split by Herschel-derived temperature, column density and stellar density. We found that the completeness-luminosity distribution doesn't vary much for any of the three physical properties. As an example, Fig \ref{f:comp_t} demonstrates the completeness-luminosity distribution for different temperature ranges, which is quite uniform. Splitting by gas column density or stellar density yields similar results. Thus, the effect of incompleteness is nearly identical for subsets of protostars located in measurably different environments. As such, variations in incompleteness do not cause systematic biases in the PLFs shown in Fig \ref{f:3range}.

%%%%%%%%%%%%%%%%%%%%%%%%%%%%%%%%%%%%%%%%%%%%%%%%%%%%%%%%%%%%%%%%%%%%%%%%%
%%%%%%%%%%%%%%%%%%%%%%%%%%%%%%%%%%%%%%%%%%%%%%%%%%%%%%%%%%%%%%%%%%%%%%%%%

\section{Discussion}
\label{s:dis}

Combining the \sf\ \fc\ observations with the \sp\ SESNA catalogue, we derived the bolometric luminosities for over 1000 protostars to build the PLF and compare sub-samples in different local environments. In this section, we link our results to existing theoretical models in order to shed light on present-day star-formation theories.

\begin{figure*}
\includegraphics[width=1\linewidth,trim={10 30 10 10}]{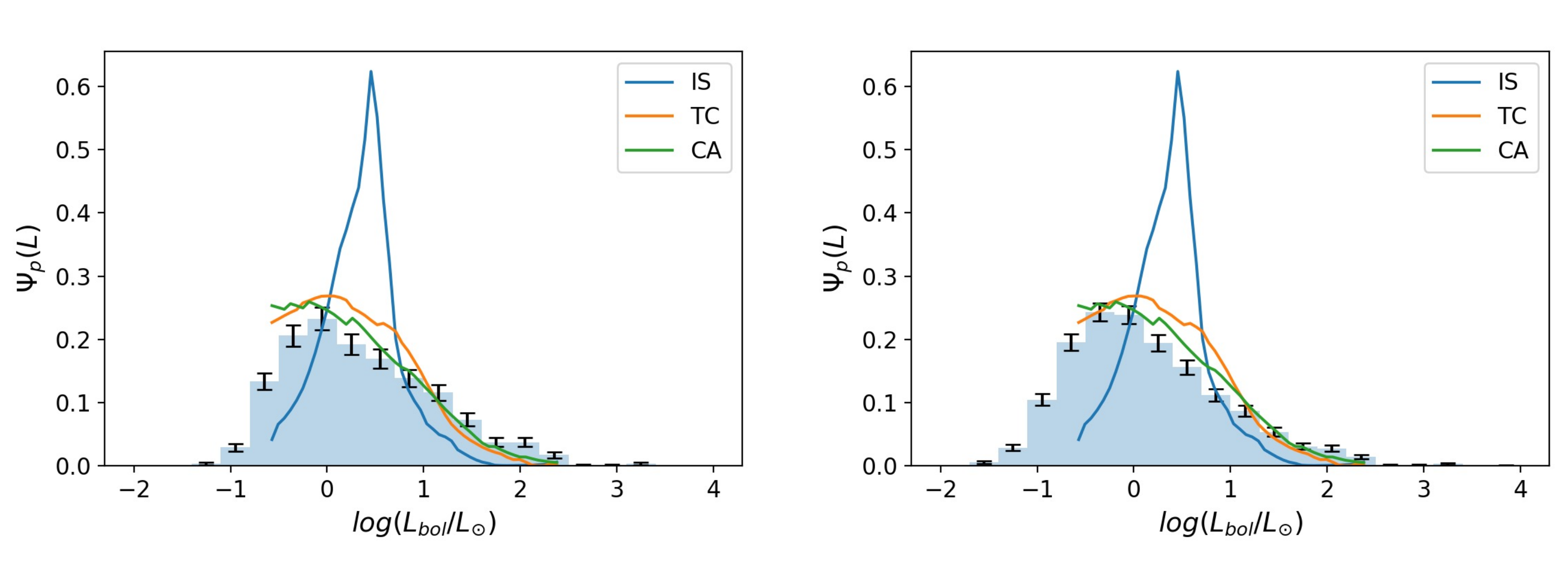}
\caption{Observed PLF overlaid with theoretical models for Sample I (left) and Sample II (right). Blue represents the isothermal sphere model, orange is the turbulent core model, and green is the competitive accretion model. The histograms are normalized according to the model range. The location of the peak is set by the mean star-formation time for the Class 0 and Class I phases, $<t_f>$, which is a function of the cloud physical conditions according to the model prediction. Here, the curves adopt the same mean time $<t_f>=0.44$ Myr as suggested by observations of local star-forming regions \citep{evans2009}.}
\label{f:model}
\end{figure*}

\subsection{Comparing with Accretion Models}
\label{ss:model}

\begin{figure*}
\includegraphics[width=0.65\linewidth]{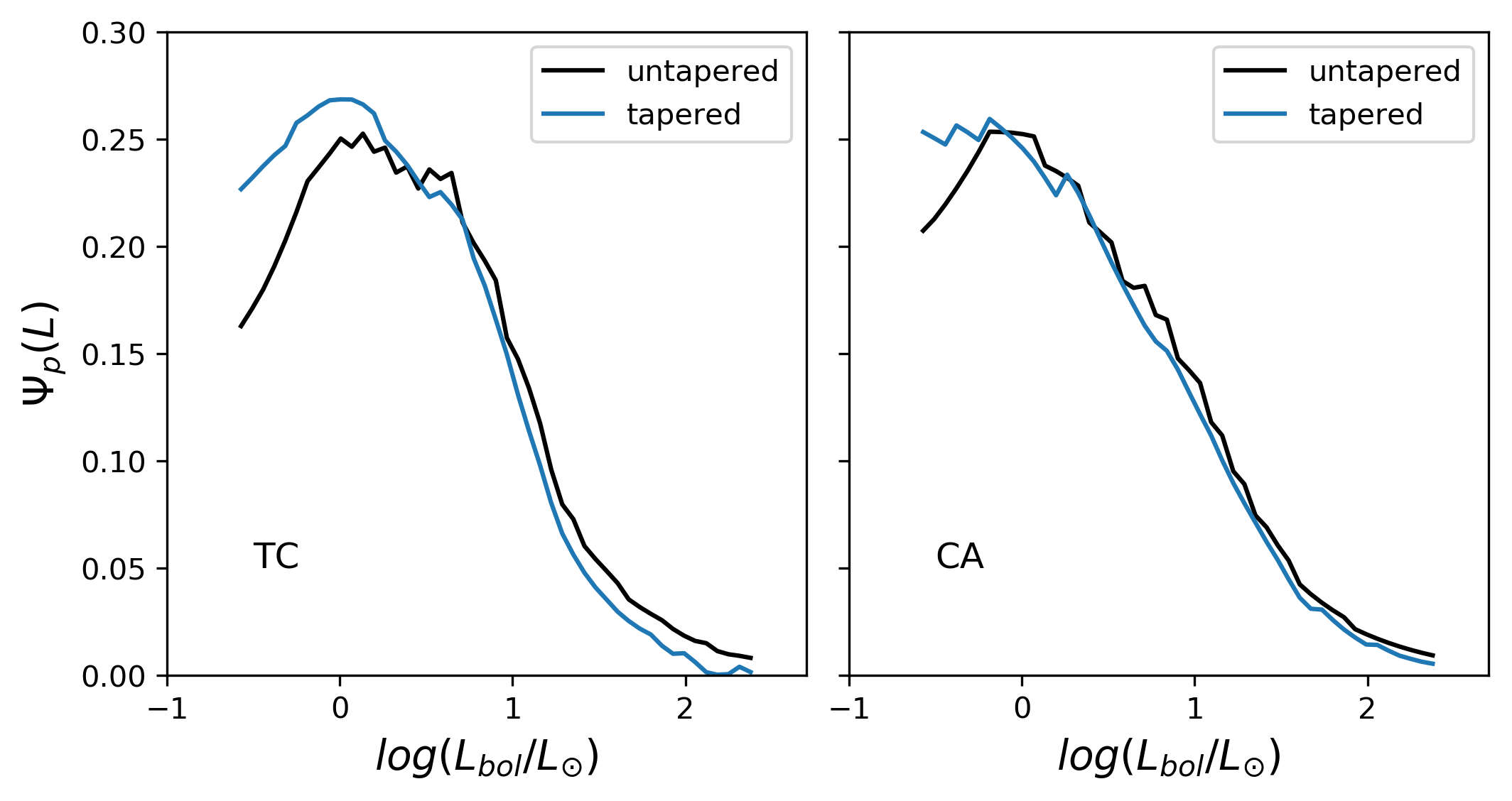}
\includegraphics[width=0.65\linewidth]{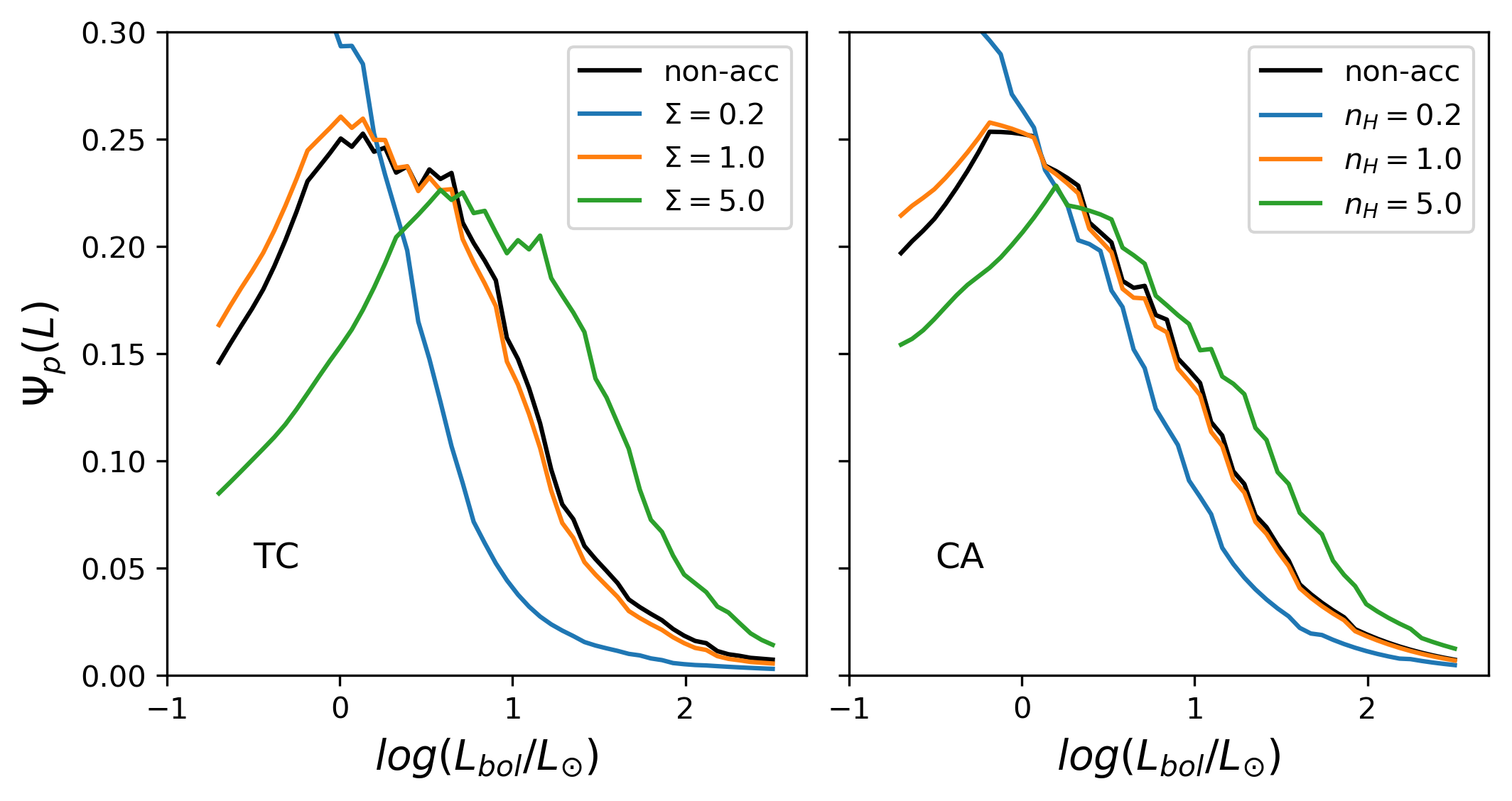}
\caption{PLFs predicted by TC and CA models with different model assumptions. The first row compares tapered and untapered models without acceleration, the second row compares accelerated models with different inputs of column/volume density. $\Sigma$ is in units of $0.1~g~cm^{-2}$ and $n_H$ is in units of $10^4~cm^{-3}$.}
\label{f:par}
\end{figure*}

The PLF can be analytically predicted by combining a protostellar accretion model and an underlying IMF, in combination with a stellar evolutionary model that provides the luminosity as a function of instantaneous and final stellar mass.  \citealt{offner2011} derived PLFs for three different accretion models: the Isothermal Sphere model (IS; \citealt{shu1977}), the Turbulent Core model (TC; \citealt{mckee2002,mckee2003}), and an approximation of the Competitive Accretion model (CA; \citealt{z1982, b1997}). The IS model is commonly used for low-mass star formation, where gas accretes from an isothermal gas sphere onto the protostar at a constant rate determined by temperature. The TC model describes high-mass star formation in which stars form from turbulently supported cores within a gravitationally bound clump of gas. In the CA model, stars accrete gas in the same gravitational potential until exhausted or ejected. The accretion rate of each protostar depends on its mass and location within the clump of gas. \citealt{mckee2010} provide analytic prescriptions based on these models and unify them into one framework. Some additional effects considered include tapering of the accretion rate in time, episodic accretion, and an accelerating star formation rate. Formulations of the TC and CA PLF models show good agreement with the PLFs obtained from hydrodynamic simulations of star-forming clouds (e.g., \citealt{Li2018, Hansen2012}). Here we take the model implementations given by \citealt{offner2011}, generate the PLF for each model and compare with our results in Section~ \ref{ss:plf}.

The analytic description for the PLF depends on several key variables, which we define here. The luminosity range depends on the lowest and highest stellar mass represented in the final IMF, $m_l$ and $m_u$, respectively. In the case of tapered accretion, the accretion rate declines prior to the end of the protostellar stage due to the shrinking core and gas dispersal \citep[e.g.,][]{offner2017}. \citet{mckee2010} parameterize tapering using a function in which accretion declines linearly with formation time, $t_f$: $\dot m = \dot m_0 (1 - (t/t_f))$, where $\dot m_0$ is a function of the instantaneous mass, $m$, and the final stellar mass, $m_f$. Here we consider accretion as a binary decision of all stars either having tapered or untapered accretion. In the case of accelerating star-formation, we follow  \citet{mckee2010} and assume star formation increases exponentially with some global  acceleration time, $\tau$: $\dot N \propto e^{(t-t_m)/\tau}$, where $t_m$ is the age of a star with mass $m$ and final mass $m_f$.
The rate of star formation should accelerate in time in a contracting gas cloud with direct evidence in a number of nearby star-forming clusters \citep{stahler2000}. The PLF normalization also depends on an accretion rate coefficient that is a function of the local environment.

For the TC model, the mass accretion rate $\dot{m}_{\rm TC}$ is proportional to the gas column density $\Sigma$ in the form of $\dot{m}_{\rm TC} \propto \Sigma^{3/4}$ (see Eq. 7 in \citealt{offner2011}). The CA model predicts a similar dependence where the accretion rate $\dot{m}_{\rm CA}$ is proportional to the gas volume density $n_{\rm H}$ ($\dot{m}_{\rm CA} \propto n_{\rm H}^{1/2}$, Eq. 22 in \citealt{mckee2010}). Presumably, a higher gas column density also indicates a higher gas volume density, but direct estimation of $n_{\rm H}$ from observations is difficult. In the IS model, the accretion rate depends only on the gas temperature, since $\dot{m}_{\rm IS} \sim c_s^3/G \propto T^{3/2}$.

The appropriate choice for the stellar mass cutoffs depends on the properties of the star-forming region. For example, a more massive star-forming region with more YSOs is statistically more likely to be forming higher mass stars. Meanwhile, the lower mass limit is sensitive to the region distance and survey resolution. To explore how the PLF shape is affected by the lower and upper mass cutoffs, we consider the simplest scenario: a PLF model that has no accretion tapering and no accelerating star formation. By taking small steps in the mass cutoffs, we confirm that within the observed luminosity range, the predicted PLFs are insensitive to the choice of $m_l$ or $m_u$ within a wide mass range ($m_l \in (3\times10^{-4},~8\times10^{-2})$, $m_u \in (5,300)$). %Really? Is this because of the normalization, i.e., the tail is longer for high mu but due to N*, the curve is normalized to ~0 at high L? In the case of $m_l$, we truncate the predicted PLF at the observational resolution limit of XX (right?), so $m_l$ has little effect on the PLF shape provided its value corresponds to luminosities are are less than those sampled by the survey. 
Thus, we fix the lower and upper cutoffs of stellar mass in the following analysis to $m_l$=0.01 and $m_u$=50. The resulting curve of each model is compared with our observed PLFs in Fig \ref{f:model}. The small modulations in the curves are produced by the stellar evolution model, which is non-monotonic and produces jumps in the protostellar radius as it passes through different stages of Deuterium burning \citep[e.g.,][]{Offner2014}. The IS model (blue) has a sharp peak between 0 and 1 on the $x$ axis, which is inconsistent with the observed data, while the TC (orange) and CA (green) models appear to describe the distribution fairly well.

Next we consider the impact of tapered accretion. As shown in Fig. \ref{f:par}, within the observed luminosity range, assuming tapered or untapered accretion rate has a small effect on the overall shape of modeled PLFs. The difference mainly lies in $log(L_{\rm bol}/L_{\odot})<0$, which is beyond the fitting range of our observed PLFs.

The PLF shape is also sensitive to the mass accretion rate. Since the mass accretion rate is proportional to the gas column density or volume density for the TC and CA models, we check the variation of the modeled PLFs by assuming different density parameters. As shown in Fig \ref{f:par}, a lower gas density leads to a lower mass accretion rate, and in turn, an excess of low-luminosity sources and deficiency at higher luminosities, although this effect becomes negligible when the density gets high enough.

%Both the TC and CA models predict an environmental dependence of the PLF. For the TC model, the mass accretion rate $\dot{m}_{TC}$ is proportional to the gas column density $\Sigma$ in the form of $\dot{m}_{TC} \propto \Sigma^{3/4}$ (see Eq. 7 in \citealt{offner2011}). The CA model predicts a similar dependence where the accretion rate $\dot{m}_{CA}$ is proportional to the gas volume density $n_{\rm H}$ ($\dot{m}_{CA} \propto n_{\rm H}^{1/2}$, Eq. 22 in \citealt{mckee2010}). Presumably, a higher gas column density also indicates a higher gas volume density, but direct estimation of $n\rm{H}$ from observations is difficult.

\begin{figure*}
\includegraphics[width=1\linewidth]{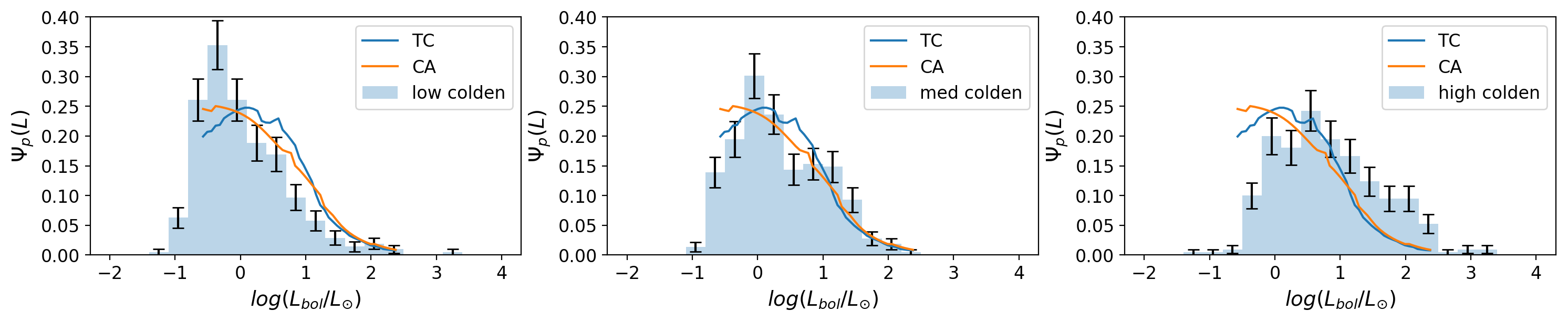}
\caption{PLFs of different gas column density ranges overlaid with untapered and non-accelerated TC and CA models for Sample I. In the labels, 'colden' is short for gas column density.}
\label{f:model_cd}
\end{figure*}

\begin{figure*}
\includegraphics[width=1\linewidth]{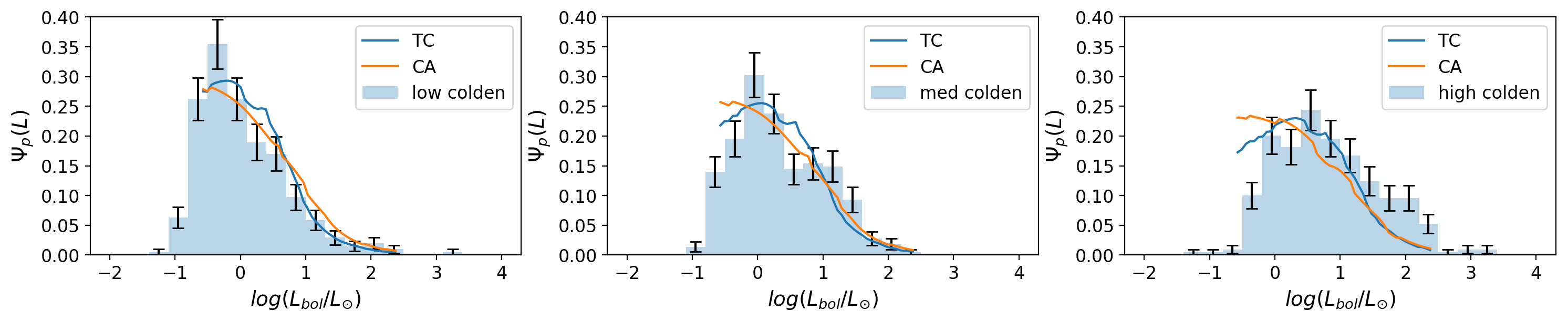}
\caption{PLFs of different gas column density ranges overlaid with untapered and accelerated TC and CA models for Sample I. Assumptions of density and mean star formation time are adjusted according to the column density ranges.}
\label{f:model_cdacc}
\end{figure*}

Fig \ref{f:model_cd} shows how the untapered and non-accelerating TC and CA models compare to our sub-samples of PLF divided by gas column density. While the PLF of the medium column density range can be described fairly well assuming the fiducial mean star formation time of $<t_f>=0.44$~Myr, both of the models tend to underestimate the fraction of low luminosity sources and slightly overestimate the fraction of high luminosity sources for the PLF of protostars forming in low-column density environments. Meanwhile, the reverse is true for the comparison of the PLF of protostars forming in the high column density environments. This discrepancy can be partly mitigated by assuming the accelerated model and adjusting the input density parameters, which suggests that the low and high density environments have different star formation times and hence effectively different mean accretion rates. Since the medium column density range is relatively well-described, we use the defaulted setting of column or volume density ($\Sigma=1$ in units of $0.1~g~cm^{-2}$ or $n_H=1$ in units of $10^4~cm^{-3}$) for this range and adjust the density parameters for other ranges according to the ratios of observed median column densities ($\Sigma=0.52$ or $n_H=0.52$ for the low column density range and $\Sigma=2.08$ or $n_H=2.08$ for the high column density range). The mean star formation time $<t_f>$ is changed accordingly with the same scale. The results are slightly improved and shown in Fig \ref{f:model_cdacc}. The models will line up even better with observations when assuming bigger changes in the density parameters, which regulates the position of the peak and the broadness of the distribution. However, the excess of observed data at the high luminosity end is still prominent in the high column density sample. There are several possible explanations for this luminosity excess. One possibility is that there are some lower mass protostars that are undergoing episodic accretion, which shifts them to higher luminosities \citep[e.g.,][]{audard2014}. Another possibility is that this feature represents an excess of intermediate mass sources. This is possible since the star formation is stochastic, and the protostars forming in a given region may not perfectly sample the (eventual) IMF. This effect should be more prominent for higher stellar masses and luminosities where the statistics are smaller. Alternatively, there could be some additional observational bias or selection effect not considered.

In summary, by comparing the PLFs predicted by different accretion models with our observed data, we can reject the IS model, but it is difficult to distinguish between the TC and CA model considering the data uncertainties. \citealt{kryu2012} found that the CA model describes the PLFs of local clouds best, but both the TC and CA models provide good agreement, which is consistent with our result. The incompleteness of the source catalogue as well as possible contamination add to the uncertainty at the lower luminosity end ($\le 1~L_{\odot}$). Both the TC and CA models depend on the environment, changing the gas densities would shift the modeled PLF. This small shift improves the fitting in the low gas column density range, but is inadequate to address the excess at the higher luminosity end in the high gas column density range. Also, these simplified accretion models do not provide a realistic description of envelope dispersal \citep{Offner2014,fischer2017}.

%Given that there is no obvious effects leading to the overestimate of high-luminosity protostars, some intrinsic adjustment to the accretion model is needed to better address the tentative environment dependence of PLFs. The differences between data and models can be partly explained by IMF sampling and episodic accretion. While the models assume the IMF is perfectly sampled, it is not the case at the high mass or high luminosity end. Also, episodic accretion is not accounted for by the smoothly varying accretion models, which could create a bump in the PLF curve or spread towards higher luminosities. Since the model fitting is always restricted to the luminosity range of observed data, these effects constrain the level of variability for the TC and CA models.

\subsection{Comparison with Previous Studies}
Whether the properties of protostars are influenced by their environment is an essential problem in star formation theories, and has been studied in nearby star-formation clouds. Previous studies have claimed that there is a dependence of protostellar luminosity on their natal environment. \citealt{kryu2012} found a significant difference between the PLFs in regions of high and low stellar density in the Orion molecular clouds, where the PLF in regions of high stellar density is biased to higher luminosities. The same dependence has been reported within \cygx\ later \citep{kryu2014}. \citealt{kryu2014} identify 2007 protostar candidates in \cygx\ with luminosities ranging from $0.1L_{\odot}$ to $3370L_{\odot}$. 
In this work, we revisit the PLFs in \cygx\ with a sample more complete at higher luminosities, investigate the dependence of PLFs not only on the stellar density, but also on the gas temperature and column density. We use the updated \sp\ SESNA catalogue with improved uncertainty and completeness products. By including \fc\ observations, we are able to improve the completeness at the high luminosity end and also get more upper limits for undetected sources. In Sample I of this work, the protostar luminosities range from $0.1L_{\odot}$ to $7610L_{\odot}$. Particularly, the luminosities of \fc\ detected sources range from $13.5L_{\odot}$ to $7610L_{\odot}$, accounting for $\sim36\%$ of sources with $L>100L_{\odot}$ in sample I.

\citealt{kryu2014} treat the incompleteness of their protostar sample in a way different from ours. They follow the approach described in \citealt{megeath2012} to determine the MIPS 24 completeness as a function of RMEDSQ, the local fluctuations due to nebulosity and neighboring point sources. Protostars with lower luminosity cannot be detected in regions with high RMEDSQ, which affects the shape of the PLF. In order to eliminate this bias, they filter out sources with either high values of RMEDSQ or low values of luminosity to achieve 90\% completeness. This results in a 'nebulosity filtered' sample of 1838 protostars, and they only use this filtered sample to compare PLFs in their analysis. 
While they use the MIPS 24 completeness only and assume a high SED slope to maximize the luminosity upper limits, we use the SESNA by-bandpass completeness products, considering all the required bandpasses and let local observing conditions set which of the multiple bandpasses is the limiter. Also, we let each protostar set the SED slope to decide the appropriate luminosity upper limits, which naturally sews in a SED slope distribution like that of the protostars themselves.
Instead of filtering sources out, we address bad completeness zones by following up with \fc\ observations to fill in the high luminosity SEDs and place better upper limits on known protostars. We confine our fitting range to $0.5<log(L_{\rm bol}/L_{\odot})<2.5$ where we have stable and high completeness, and extend the range once applying our completeness correction. Also, we have shown that the effect of incompleteness is nearly identical for protostars in different local environments.

\citealt{kryu2014} account for three types of contamination in their sample, including edge-on disks, highly reddened Class II sources and galaxies. For each type of contaminant, they predict the luminosities for ensembles of contaminating sources as generated by Monte Carlo simulations, and then remove sources from the PLF which has the closest luminosity to the contaminant. They conclude that the contaminants account for only 14.2\% of the total protostar sample, among which 6.7\% comes from edge-on disks. In this work, we confirm that the contamination from galaxies or reddened Class IIs is negligible in our sample, the edge-on disks dominate the protostellar contamination rate with an upper limit of $\sim$7.7\% and strongly bias to low extinctions.

As shown in Fig \ref{f:3range} and Table \ref{tab:3range}, there is no clear trend of PLF slopes according to the gas temperature or gas column density, but PLFs in regions of higher stellar density tend to have flatter slopes, which is consistent with the previous result. The differences of PLF slopes between low and high stellar density regions are around $3\sigma$ for both Sample I and Sample II, while conducting the completeness correction enlarges this discrepancy to $\sim 4\sigma$. PLFs in regions of different gas column densities show similar scale of differences, but the steepest slope appears at the medium column density range. Thus, there is a weak dependence of PLFs on the surrounding gas column density and stellar density, but considering the data uncertainties, the variations cannot be used to announce a conclusive trend. Meanwhile, the variation of PLF slopes in regions of different temperatures is as small as $\sim 1 \sigma$, which means that the PLF shape is almost identical at different temperatures. The mild difference between our result and previous work might arise mainly from the sample selection. Different approaches of contamination correction for reddened Class II sources, edge-on disk sources and galaxies lead to different samples of protostars and also different ranges of stellar densities. Also, adopting different luminosity ranges for fitting can slightly affect the PLF slopes.
Another difference arises from the method of comparing PLFs. While we compare the best-fit slopes between subsets of PLFs, \citealt{kryu2014} used the KS test to show the statistical differences.
%Several studies have shown the gas column density is correlated with the clustering of young stars. For instance, the stellar and gas volume density in the central $\sim0.5\ pc$ of the Orion Nebula Cluster are correlated, which provides strong evidence that the protocluster structure is regulated by the gas filament \citep{stutz2018}. \citealt{pokhrel2020} found the same correlation in 12 nearby molecular clouds, with the stellar density following gas column density squared. Meanwhile, \citealt{pokhrel2021} reported a constant star formation efficiency per freefall time across a wide range of adopted column density thresholds. Therefore, the relation between gas column density and local stellar density is complicated, it is hard to tell whether there is an intrinsic dependence of protostellar luminosity on either side.

\subsection{Implications for the Star-formation Process}

Within the \cygx\ complex, several different star-forming regions show different distributions of gas, stars and nebulosity. The variation of these morphologies indicates that both radiatively triggered and spontaneous modes of star formation are operating in \cygx, adding to the complexity of PLFs in this region. The luminosity of protostars is a combination of intrinsic stellar luminosity and accretion luminosity, so the differences in the observed PLFs could not only be attributed to the different final masses of the stars, because the IMF may vary in different cloud regions, also to the differences in the accretion rate. Moreover, since stars usually form in clusters, the feedback from newborn stars could have profound effects on other stars as well as their surrounding gas. For instance, a newborn protostar may heat the gas around existing stars, make the natal environment warmer and impede fragmentation. \citealt{Li2018} and \citealt{Hansen2012} investigate the PLF with simulations of radiation feedback and outflows and find good agreement with the observed PLF. However, how the shape of PLF depends on physical parameters of the cloud is still poorly understood.

We have found a tentative dependence of PLFs on the local stellar density and gas column density, where the PLF in regions of high stellar density or high gas column density is biased to the high luminosity end. This could be explained by primordial mass segregation, with more massive stars tending to form in denser regions. So far, there is no strong evidence for the environmental dependence of characteristic stellar mass \citep{offner2014,Lee2020}. Thus, it is possible that the observed dependence of PLFs on density and clustering is relatively weak. Apart from mass segregation, the accretion processes provide another approach to address the PLF differences. The TC model predicts an increasing mass accretion rate with increasing gas column density. Similarly, the CA model claims that protostars in dense, clustered environments tend to have high mass accretion rate, especially for the most massive stars. In principle, both TC and CA models predict the PLF variation seen in our observations, although the shapes are not a perfect match to the observed PLFs.

Another complexity arises from the relationship between stellar density and gas density, which is predicted by the CA model. Some turbulent hydrodynamic simulations of molecular gas with self-gravity and radiative feedback can also reproduce this trend \citep{pokhrel2020}. In \cygx, regions with high stellar density tend to contain high column density of molecular gas, and in turn lead to higher star formation efficiencies \citep{krum2010, pokhrel2021}. It is difficult to distinguish between the effects of stellar density and gas column density on the PLF differences.

We do not see the PLFs varying much in regions with different gas temperatures. The temperature of parent clouds could affect the core temperature of newborn protostars. However, as the protostars evolve to the IR-emitting stage, the observed luminosities show little dependence on the environment temperature, and thus the PLFs are in general consistent within the whole temperature range. This further rules out the IS model, in which the only variable is the temperature. However, bringing in episodic accretion to the IS model could produce a better fit to the data (e.g., \citealt{dunham2014}).
%%%%%%%%%%%%%%%%%%%%%%%%%%%%%%%%%%%%%%%%%%%%%%%%%%%%%%%%%%%%%%%%%%%%%%%%%%%%%%

\section{Conclusions}
\label{s:con}
We test and adopt an empirical correlation between $L_{bol}$, $L_{MIR}$ and $\alpha$ to conduct a systematic analysis of the protostellar luminosity function in \cygx, based mainly on \sf\ \fc\ observations and the existing \sp\ SESNA catalogue. Our main results and conclusions are as follows:

\begin{itemize}

\item[1.]
To get the SED slope $\alpha$, we adopt \sp\ IRAC 4 bands + MIPS 24 band to do the SED fitting, and use the \fc\ 31~$\mu$m photometry to extrapolate the 24~$\mu$m flux when necessary. Through the method described in \citealt{kryu2012}, we derive the bolometric luminosities for over 1000 protostars in \cygx, while possible biases arising from missing bands are addressed. Making use of the SESNA completeness map, \fc\ flux and uncertainty map, as well as the WISE data, we further derive the upper limits of $\alpha$ and $L_{\rm bol}$ for $\sim$700 protostars with incomplete photometry.

\item[2.]
We plot the distribution of protostellar luminosities according to several selected protostar samples and conduct power-law fitting to characterize the PLF shape. Given the big incompleteness of measurement at the low luminosity end, we choose a fitting range of $0.5<log(L_{\rm bol}/L_{\odot})<2.5$. The PLFs are well described by a simple power law with an index of $\sim-0.5$. Possible contamination from edge-on disks and galaxies shows little effect on the fitting results.

\item[3.]
We further use the Herschel temperature and column density map in \cygx\ to compare the PLFs in different local environments in terms of gas temperature, gas column density and stellar density. We find no obvious dependence of PLFs on the gas temperature, but there is evidence that the PLFs in regions of high stellar density or high gas column density exhibit some excess at higher luminosities. This dependence can be explained by primordial mass segregation and/or the environmental dependence of the mass accretion rate, as predicted by the Turbulent Core (TC) and the Competitive Accretion (CA) models.

\item[4.]
We compare our observed PLFs in \cygx\ with accretion models. While the IS model is strongly disfavored, both the TC and CA models are generally consistent with our results, and it is hard to evaluate these two models given the available statistics. The TC and CA models predict increasing mass accretion rates with increasing gas density and thus an excess of high luminosity protostars in the PLF. However, current models are unable to fit the observed PLFs over the full range of observed luminosities.

\end{itemize}

The upcoming JWST instruments and TolTEC camera on the LMT will provide more reliable and complete protostar photometries and improve the SED sampling at both mid-IR and millimeter wavelengths. The Mid-Infrared Instrument (MIRI) on-board the JWST allows detection of protostars in more distant regions out to 2 kpc and beyond, such as some infrared dark clouds (IRDCs). These additional data together with better completeness for lower luminosity SEDs will provide better constraints on star formation theories.

%%%%%%%%%%%%%%%%%%%%%%%%%%%%%%%%%%%%%%%%%%%%%%%%%%%%%%%%%%%%%%%%%%%%%%%%%%%%%%

\section*{Acknowledgments}

This work is based in part on observations made with the NASA/DLR Stratospheric Observatory for Infrared Astronomy (SOFIA). SOFIA is jointly operated by the Universities Space Research Association, Inc. (USRA), under NASA contract NNA17BF53C, and the Deutsches SOFIA Institute (DSI) under DLR contract 50 OK 0901 to the University of Stuttgart.

Financial support for this work was provided by NASA through award \#05-0181, \#07-0225, and \#08-0181 issued by USRA. This work is based in part on archival data obtained with the Spitzer Space Telescope, which was operated by the Jet Propulsion Laboratory, California Institute of Technology under a contract with NASA. Support for this work was provided by NASA ADAP grants NNX11AD14G, NNX13AF08G, NNX15AF05G, and NNX17AF24G.

We thank the referees for their helpful comments that improved the clarity of our presentation.

\section*{DATA  AVAILABILITY}

The data described in \S~\ref{s:obs} are available in the data archives accessible via SOFIA database ($https://irsa.ipac.caltech.edu/Missions/sofia.html$). Processed data products underlying this article will be shared on reasonable request to the authors.

\appendix
\section{Edge-on Disk Contamination}
\label{a:1}

\citealt{guter2009} estimated an upper limit to edge-on disk contamination to their protostellar tallies in their Spitzer survey of young clusters by simply invoking evidence of substantial stellar feedback around some young clusters to argue that those are no longer actively forming new stars. Where that is the case, any protostars identified within these regions would be spurious classifications for stars with edge-on disks. Two clusters were selected for their strong evidence of feedback, low gas column densities, and yet reasonably rich populations of stars, both resulting in contamination estimates of 3.5\% of disks that could be misidentified as protostars. Given the assumption of truncated star formation, this value is an upper limit on the number of disks misclassified as protostars under the SESNA classification scheme. We can apply the same sort of logic to the SESNA \cygx\ catalog, carving out the regions of low gas column density, computing the protostar to disk ratio, and assuming that all such protostars are actually edge-on disks. The resultant maximal edge-on disk contamination rate is 3.9\% , extremely similar to the value given by \citealt{guter2009}.

However, in the case of the present analysis, we have the added requirement of 24~$\mu$m photometry. As noted above, this reduces the number of YSOs considered, and precipitously so in the case of stars with disks owing to their declining mid-IR spectral energy distribution. If we assume that all disks regardless of inclination are similarly filtered by the 24~$\mu$m requirement, the greater loss of disks relative to protostars would lead us to the conclusion that the edge-on disk contamination rate must decline by at least a factor of two relative to the above estimates, namely $\sim$2\%.

\begin{figure}
\includegraphics[width=1\linewidth]{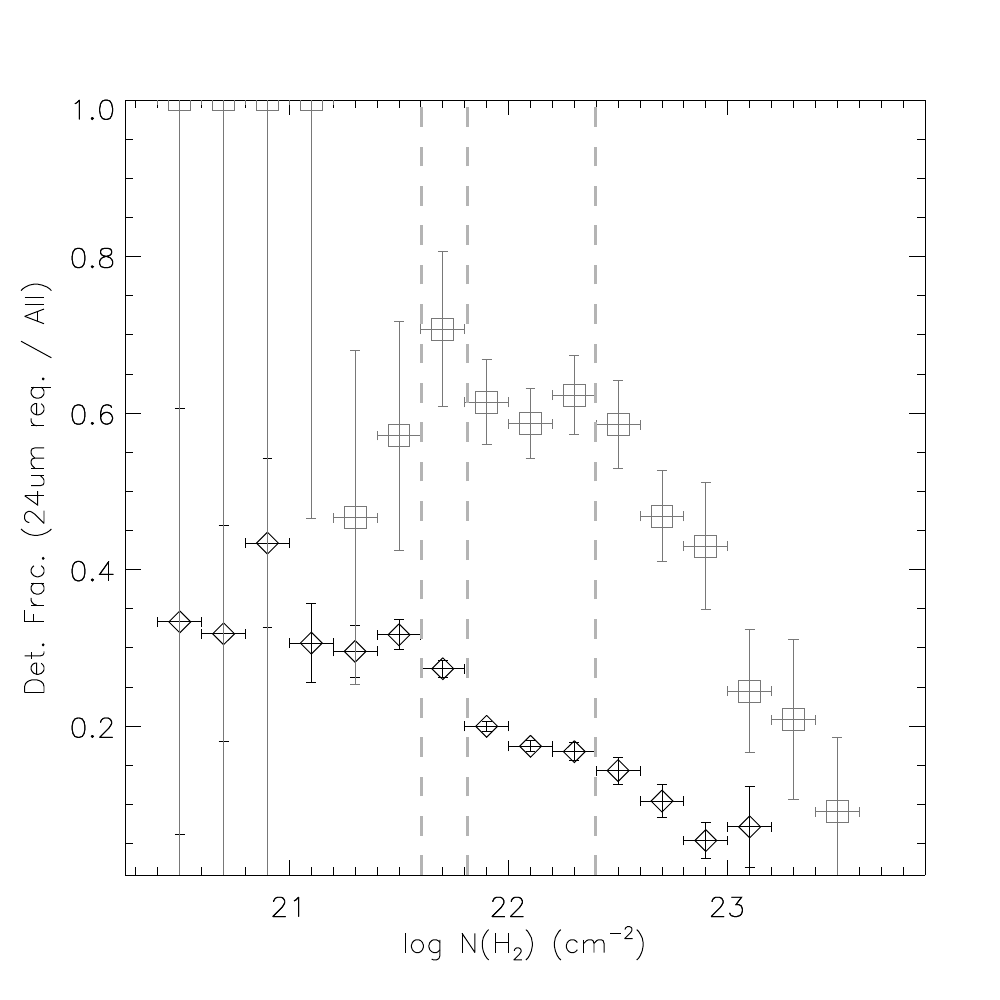}
\caption{24~$\mu$m detection fractions for SESNA YSOs in \cygx\ binned by Herschel-derived gas column density. Gray squares mark the protostar
measurements, and the black diamonds mark the stars with disks. Gray dashed vertical lines denote frequently used column density separations in our analysis.}
\label{f:edgeon}
\end{figure}

As can be seen in Fig.~\ref{f:edgeon}, the 24~$\mu$m detection fraction for disks is relatively flat ($\sim$32\%) at low column densities with $log N(H_2) < 21.6$. Then it declines by a factor of two at higher densities ($21.8 < log(NH_2) < 22.4$). In contrast, the protostar detection rates are consistent with a flat and relatively high fraction ($\sim$60\%) throughout this range. The relative lack of correlation among the detection fractions offers us additional confidence that the edge-on disk contamination rate is rather low.

Regardless, one can attempt a direct estimate of the edge-on contamination by assuming that the 24~$\mu$m detection fraction is underestimated for protostars at low column densities by virtue of the presence of the low-fraction contaminant, edge-on disks. In this model, we assume that the true protostellar detection fraction is flat at $\sim$69\% as indicated by the measurement in Fig.~\ref{f:edgeon} at $21.6 < log N(H_2) < 21.8$, and the lower mean detection fraction of 60\% at lower column densities is set by an unknown fraction of disks with detection fraction 32\%. Then the fraction of contaminants in the protostar sample is solved as 25\%. Multiplying this fraction by the protostar-to-disk ratio of 7.7\% in the 24~$\mu$m detected low column density sample yields an edge-on disk contamination rate of 1.9\%, but with sufficient uncertainty to encompass 0-3.5\%.

To sum up, the maximal edge-on disk contamination rate is estimated as 3.9\%, which can be used to test the maximal effect of this contamination on our PLFs and their fitting results. %As a comparison, \citealt{kryu2014} report a contamination rate of 14.2\% within their total protostar sample, among which $\sim$47\% comes from edge-on disks.

\section{Reddened Class II Contamination}
\label{a:2}
\citet{kryu2012} and \citet{kryu2014} estimated that there was substantial contamination in their protostar sample from Class II YSOs that are reddened by being embedded in or simply located behind dense molecular gas structures that are also rich with cold dust. While the SESNA source classification process was built to be more reddening safe, no approach is perfect. Here we perform a simple experiment to estimate reddened Class II contamination in our protostar samples.

We select all SESNA-classified Class IIs and Transition Disks (Class II and
II*, respectively) that have MIPS 24~$\mu$m photometry with $\sigma([24])<0.2$~mag and an estimate of line of sight extinction in
the range $0<A_K<1$~(mags) as measured from their 1-5~$\mu$m photometry
\citep{guter2009}. This results in a sample of 2380 sources with a median $A_K$ of $0.51$~mag.

Based on the Herschel-derived column density value ranges adopted for
our low, medium, and high column density protostar samples (see Table
\ref{tab:3range}), and assuming $N(H_2) = 10^{22}~cm^{-2} \approx 1~A_K$, our typical column density for the protostars in each sample corresponds to 0.5, 1.0, and 2.0~$A_K$, respectively. Given the agreement between the typical extinction values for the low-extinction Class II sample and the low column density protostar sample, we assume the low column density sample has no reddened class II contamination and/or they would be indistinguishable from the edge-on disk contamination characterized in Appendix \ref{a:1}.

To estimate the potential protostar contamination in the medium and high
column density samples, we artificially apply additional reddening of 0.5 and 1.5 mags $A_K$ to the low extinction Class II sample's photometry using
the reddening and extinction laws \citep{flaherty2007} and reclassify
the sources with the SESNA scheme. No attempts were made to reduce the
signal-to-noise ratio of the adjusted photometry nor to delete any photometry that was sufficiently extinguished such that the source would have gone undetected in some or all of the observations. Both of these treatments would have the effect of reducing the estimated protostar contamination.

For 0.5 and 1.5 mags $A_K$ added, the false protostar counts are 24 and
69, with returned Class II counts of 2317 and 2183, for reddened Class
II contamination rates of 1.0\% and 3.2\%, respectively. Applying these
rates to the full sample of Class IIs found within the sample column
density ranges as our protostar divisions (i.e. 2608, 681, and 347 for
low to high column density), we estimate 0, 7, and 11 reddened Class II
contaminants in the low to high column density samples. Those samples
are all 300-400 sources in size, thus the protostellar contamination
rate is less than 4\% in all of them. Furthermore, contaminating sources
are somewhat extinguished and thus are prone to skew toward relatively
low measured luminosities.

\end{document}